\documentclass[10pt, conference]{IEEEtran}
\usepackage{amsmath,graphicx,cite}
\usepackage{amsfonts}
\usepackage{algpseudocode}
\usepackage[ruled,linesnumbered]{algorithm2e}
\usepackage{verbatim}
\usepackage{longtable}
\usepackage{rotating}
\usepackage{float}
\usepackage{multirow}
\usepackage{tabularx}
\usepackage{bigints}
\usepackage{tikz}
\usepackage{multirow} 
\usepackage{amsmath,amstext,amsfonts,amssymb}
\usepackage{amsbsy}
\usepackage{amsthm}
\usepackage{diagbox}
\usepackage{mathabx}
\usepackage{mathrsfs}
\usepackage{bbm}
\usepackage{color}
\usepackage{caption}
\usepackage{subcaption}
\usepackage{morefloats}
\usepackage{float}
\usetikzlibrary{arrows,shapes,positioning,shadows,trees}
\usetikzlibrary{decorations.pathreplacing}
\newcommand{\mb}{\mathbf}

\newcommand{\mc}{\mathcal}

\newtheorem{theorem}{Theorem}[section]
\newtheorem{definition}[theorem]{Definition}

\newtheorem{proposition}[theorem]{Proposition}

\title{Goal-Oriented Quantization:\\Analysis, Design, and Application to Resource Allocation}
\author{
        Hang~Zou,~\IEEEmembership{Student~Member,~IEEE,}
        Chao~Zhang,~\IEEEmembership{Member,~IEEE,}
        Samson~Lasaulce,~\IEEEmembership{Member,~IEEE,}
        Lucas~Saludjian,~
        and~Vincent~Poor~\IEEEmembership{Fellow,~IEEE,}
        
\thanks{H. Zou, S. Lasaulce and C. Zhang was with Laboratory of Signals and Systems (L2S), CentraleSupelec, University of Paris Saclay, Gif-sur-Yvette, France, e-mail:  }

\thanks{This work has been funded by the RTE-CentraleSupelec Chair on ``The Digital Transformation of Electricity Networks".}}
\begin{document}
%
\maketitle
%
%

\begin{abstract}
In this paper, the situation in which a receiver has to execute a task from a quantized version of the information source of interest is considered. The task is modeled by the minimization problem of a general goal function $f(x;g)$ for which the decision $x$ has to be taken from a quantized version of the parameters $g$. This problem is relevant in many applications e.g., for radio resource allocation (RA), high spectral efficiency communications, controlled systems, or data clustering in the smart grid. By resorting to high resolution (HR) analysis, it is shown how to design a quantizer that minimizes the gap between the minimum of $f$ (which would be reached by knowing $g$ perfectly) and what is effectively reached with a quantized $g$. The conducted formal analysis both provides quantization strategies in the HR regime and insights for the general regime and allows a practical algorithm to be designed. The analysis also allows one to provide some elements to the new and fundamental problem of the relationship between the goal function regularity properties and the hardness to quantize its parameters. The derived results are discussed and supported by a rich numerical performance analysis in which known RA goal functions are studied and allows one to exhibit very significant improvements by tailoring the quantization operation to the final task. 
\end{abstract}

\section{Introduction}
\label{sec:introduction}
 
Since the pioneering and fundamental works of Shannon \cite{seminal-Shannon}, the dominant paradigm for designing a communication system is that communications must satisfy quality requirements. Typically, the bit error rate, the packet error rate, the outage probability, or the distortion level must be minimized. It turns out that the conventional paradigm consisting in pursuing communication reliability or possibly security may not be suited to scenarios such as systems where communications occur in order for a given task to be executed. For instance, transmitting an image of 1 Mbyte to a receiver that only needs to decide about the absence/presence of a given object in the image might be very inefficient. In this example, the receiver only needs one bit of information and this bit could have been directly sent by the transmitter and make the use of the communication and computation resources much more efficient. This simple example shows the potential of making a communication task- or goal-oriented (GO).

In this paper, the focus is on the problem of signal compression when the compressed signal is used for a given task which is known. More precisely, we focus on the signal quantization problem, which is often a key element of a signal transmitter. \textcolor{black}{Introducing and developing a goal-oriented quantization (GOQ) approach} is very relevant for many applications. We will mention three of them. First, it appears in controlled networks that are built on a communication network. A simple example is given by modern power systems such as the smart grid. A data measurement system such as a smart meter may have to quantize or cluster the measured series for complexity or privacy reasons \cite{poor-privacy}. It is essential that the quantization or clustering operation does not impact too much the quality of the decision (e.g., a power consumption scheduling strategy) taken e.g., by an aggregator. Second, GOQ is fully relevant for wireless RA problems. For instance, if a wireless transmitter receives some quantized information from the receivers/sensors through a limited-rate feedback channel \cite{zheng-TSP-2007,Lee-TWC-2015,Yeung-TC-2009,Love-JSAC-2008,Kountouris-ICASSP-2007}. Third, for future wireless communication systems such as 6G systems \cite{Saad-NW-2020,Giordani-CM-2020,Bertin-IEEE-2022,Letaief-CM-2019}, GOQ and more generally GO data compression constitutes a very powerful degree of freedom of increasing final spectral efficiency since only the minimum number of bits to execute the task is transmitted through the radio channel. 

\begin{figure}[tbp]
\centering{}\includegraphics[scale=0.16]{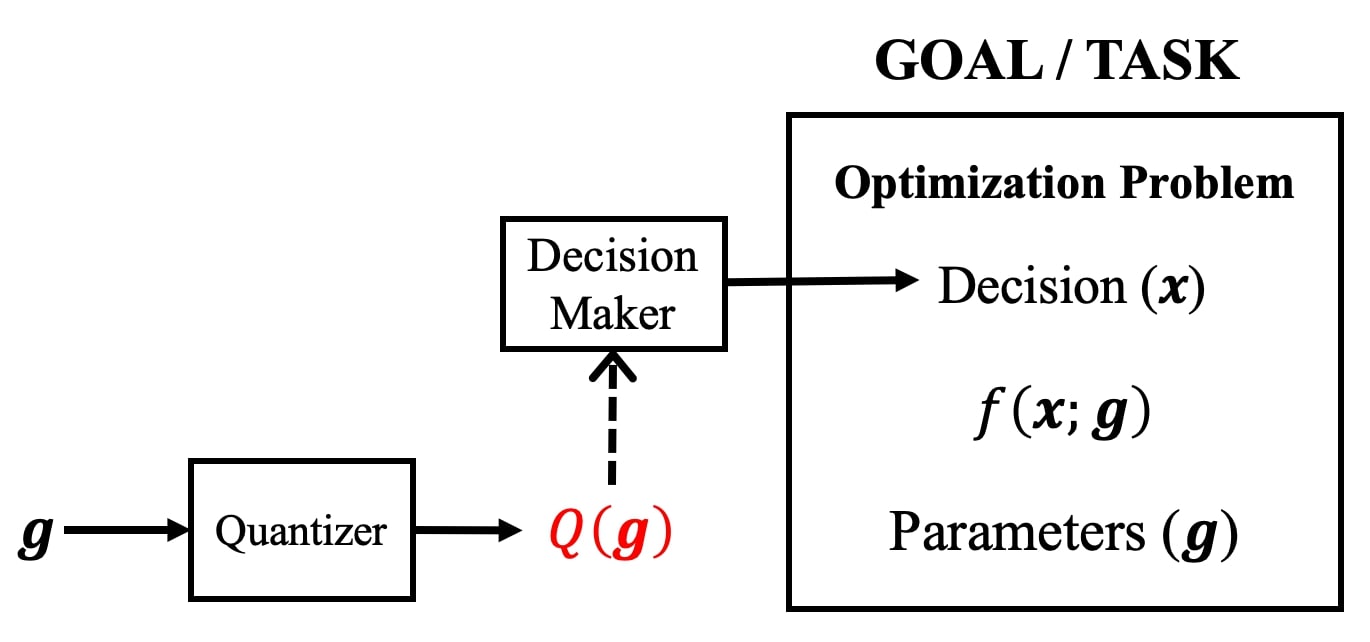}
\caption{Proposed definition for the goal-oriented quantization \textcolor{black}{approach} \label{fig:GOQ-OP}}
\end{figure}

The conventional quantization \textcolor{black}{approach} \cite{Gray_TIT_1998} is to minimize some distortion measure between the original signal and its representation, regardless of the system task. In the literature, there exist works on the problem of adapting the quantizer to the objective. For instance, in the wireless literature, the problem of quantizing channel state information (CSI) for the feedback channel has been well studied (see e.g., \cite{Rao-TIT-2006} for a typical example). The practical relevance of low-rate scalar quantizers to transmit high dimensional signals has been defended for MIMO systems in  \cite{Rini-2017}\cite{Choi-2017}\cite{Li-2017}. By combining the system task with the quantization process, \cite{Eldar-TSP1-2019}\cite{Eldar-ISIT-2019} investigated the influence of scalar quantization on specific tasks and characterized the limiting performance in the case of recovering a lower dimensional linear transformation of the analog signal and reconstruction of quadratic function of received signals. \textcolor{black}{Deep-learning-based quantizers have also be considered in \cite{Choi-arxiv-2019,Hanna-JSAIT-2020,Hanna-JSAIT-2021,Sohrabi-TWC-2012} to adapt to the task by training neural networks.} The main point to be noticed is that for all existing works either the impact of quantization on a given performance metric is studied or a very specific performance metric is considered (the Shannon transmission rate being by far the most popular metric) and the proposed quantizer design is often an ad hoc scheme. In contrast with this line of research works, we introduce a general framework for GOQ \textcolor{black}{illustrated in Fig. \ref{fig:GOQ-OP}}. The task or goal of the receiver is chosen to be modeled by a generic optimization problem (OP) which contains both decision variables and parameters. One fundamental point of the conducted analysis is that both for the performance analysis and the design, the goal function is a generic function $f(x;g)$, $x$ being the decision \textcolor{black}{with dimension $d$}  to be made based on a quantized version of the function parameters $g$ \textcolor{black}{with dimension $p$}. \textcolor{black}{This setting allows us to derive analytical results and acquire completely new insights} into how to adapt a quantizer to the goal, these insights relying in part on the high resolution (HR) regime analysis \cite{Misra_TIT_2011,Fleicher_TIT_1964,Farias-TSP-2014}. 

To be sufficiently complete concerning the technical background associated with the present contributions, we also would like to clearly position our works w.r.t. recent works on semantic communications 
\cite{Kountouris-CM-2021,Shi-CM-2021,Barbarossa-CN-2021,Zhang-ISIT-2021,Qin-TSP-2021,Sana-2021,Qin-JSAC-2021,Qin-JSAC2-2021,Yun-ISWCS-2021,Saad-arXiv-2022,Niyato-arXiv-2022,Lan-2021,Debbah-Tcom-2021}. Semantics is employed here with its etymological meaning, that of significance. It can be seen as a measure of the usefulness/importance of messages with respect to the system task \cite{Kountouris-CM-2021}. There have been several tutorials and surveys to discuss possible structures and architectures of this novel communication paradigm. By studying the semantic encoder and semantic noise, \cite{Shi-CM-2021} proposed two models based on shared knowledge graph and semantic entropy, respectively. Reference \cite{Barbarossa-CN-2021} indicated that by properly recognizing and extracting the relevant information to the system task, the communication efficiency and reliability can be enhanced without using more bandwidth. In \cite{Kountouris-CM-2021}, it is explained how semantic information attributes of transmitted messages could be exploited,  which entails a task-oriented unification of information generation, transmission, and reconstruction. By introducing intrinsic states and extrinsic observations, \cite{Zhang-ISIT-2021} uses indirect rate-distortion theory to characterize the reconstruction error of semantic information induced by lossy source coding schemes. Information bottleneck is also an approach to find the optimal tradeoff between compressing and reliability. \textcolor{black}{Inspired by this approach, \cite{Qin-TSP-2021} proposed a relevant loss function whose relevance was supported in \cite{Sana-2021} and designed an end-to-end DeepSC network architecture, using Transformer as the semantic encoder and joint source-channel coding schemes to ensure the semantic information transmission. Similar models \cite{Qin-JSAC-2021}\cite{Qin-JSAC2-2021} are extended to audio transmission and Internet-of-things (IoT) applications.} Other learning tools have also been implemented to extract important attributes in semantic communications, such as reinforcement learning\cite{Yun-ISWCS-2021}, curriculum learning \cite{Saad-arXiv-2022}, and distributed learning \cite{Niyato-arXiv-2022}\cite{Lan-2021}. Some additional information can also be used for the semantic encoder, such as contextual reasoning\cite{Debbah-Tcom-2021}. Compared to the quoted works, three main points have to be noticed. First, most works focus on the novel communication architecture or use learning tools to extract important features but the works are not supported by theoretical derivations. Second, we not only consider the transmission problem of the semantic information but also the influence of distorted information on the subsequent decision-making (DM) entity and the system task, namely, how the semantic information exchange will affect the system performance (effectiveness level). Third, we address a precise technical problem which is the quantization problem and assume a fully generic goal. The closest contributions to the present work have been produced by the authors through \cite{Zhang_Wiopt_2017}\cite{Zou_WinCom_2018}\cite{hang-pimrc-2019}\cite{Zhang-AE-2021}. To the best of the authors knowledge, the concept of GOQ has been introduced for the first time in \cite{Zhang_Wiopt_2017} and applied in other contexts in \cite{Zou_WinCom_2018}\cite{hang-pimrc-2019}\cite{Zhang-AE-2021}. In these references, mainly numerical results are provided and the focus is on a Lloyd-Max (LM)-type algorithm \cite{Lloyd}\cite{Max}. In particular the formal HR analysis is not conducted and the fundamental role of the goal function is not investigated.

This paper is structured as follows. In Sec. \ref{sec:problem_formulation}, we define the performance metric of a GO quantizer. In Sec. \ref{sec:scalar_approximation}, the performance analysis of scalar GOQ is conducted in the HR regime and the impact of the goal function on the optimality loss  \textcolor{black}{(OL)} is assessed through analytical arguments. In Sec. \ref{sec:vector_approximation}, we address the more challenging case of vector GOQ by providing an HR equivalent of the HR OL and a practical GOQ algorithm. In Sec. \ref{sec:Numerical_Results}, we show the potential benefit from using GOQ for important RA problems that are relevant for quantizing information in wireless, controlled, and power systems. Sec. \ref{sec:Conclusions} concludes the paper.

\section{Problem Formulation}
\label{sec:problem_formulation}

\begin{definition} \textcolor{black}{Let $d\geq 1$ be an integer and $\mathcal{G}$ be a subset of $\mathbb{R}^d$.} Let $M \geq 1$ be an integer. An $M-$quantizer $\mathcal{Q}_M$ is fully determined by a piecewise constant function $Q_M: \mathcal{G} \rightarrow \mathcal{G} $ that is defined by $Q_M(g) = z_m$ for all $z_m \in \mathcal{G}_m$ where: $m \in \{1,...,M\}$, the sets $\mathcal{G}_1,...,\mathcal{G}_M$ are called the quantization regions and define a partition of $\mc{G}$, and the points $z_1,...,z_M$ are called the region representatives. 
\end{definition}

Since $M$ is a fixed number, from now on and for the sake of clarity, we will omit the subscript $M$ from the quantization function and merely refer to it as $Q$. We will only make $M$ appear for comparison purposes, mainly in the simulations. Also, when needed, we will also use the quantity $R = \log_2 M$ which represents the number of quantization bits per sample. \textcolor{black}{Equipped with these notations, we can now define mathematically the GO approach we propose for quantization.}

\begin{definition} Let $\chi(g)$ be the decision function providing the minimum points for the goal function $f(x;g)$, \textcolor{black}{whose decision variable is $x \in \mathbb{R}^p$ ($p\geq 1$ is an integer)}, $g$ being fixed: 
\begin{equation}
    \chi(g)\in\arg\underset{{x}\in\mathcal{X}}{\min} \quad f({x};{g})
    \label{eq:ODF}.
\end{equation}
The optimality loss induced by quantization is defined by:
\begin{equation} 
    \textcolor{black}{L} \left( \textcolor{black}{Q};f\right) = \alpha_f
    \int_{{g}\in\mathcal{G}} \left[f \left(\chi \left({Q} \left( {g}\right)\right);{g} \right) - f \left(\chi \left({g} \right);{g} \right) \right] \phi \left({g} \right)\mathrm{d}{g}
    \label{eq:def-OL}
\end{equation}
where $\phi$ is the probability density function (p.d.f) of $g$ and $\alpha_f >0$ is a scaling/normalizing factor which does not depend on $Q$.
\end{definition}
 
Several comments concerning the OL definition are in order. Note that the conventional quantization \textcolor{black}{approach} can be obtained from the GOQ \textcolor{black}{approach} by observing that \textcolor{black}{the second term of the OL functional $L(Q;f)$ (that is, a function of function)} is independent of $Q$ and by specializing $f$ as $f(x;g) = \| x - g \|^2$, $\|. \|$ standing for the Euclidean norm. With the conventional \textcolor{black}{approach}, quantization aims at providing a version of $g$ that resembles to $g$. However, under the GOQ \textcolor{black}{approach}, what matters is the quality of the end decision taken. The design of such a quantizer therefore depends on the mathematical properties of $f$ and the underlying decision function $\chi$, which constitutes a key difference w.r.t. the conventional \textcolor{black}{approach}. In this respect, studying analytically the relationship between the nature of $f$ and the quantization performance is a nontrivial problem. For instance, for a fixed OL level, how do the functions requiring a small (resp. large) $M$ (that is, a small -resp. large- amount of quantization resources) look like? \textcolor{black}{The normalizing factor $\alpha_f$ is precisely introduced to conduct fair comparisons between different goal functions.} From the OL definition, it can also be noticed that the knowledge of the p.d.f. of $g$ is implicitly assumed. One may replace the statistical mean with an empirical mean version and rewrite the OL under a \textcolor{black}{data-based form where the integral is replaced with a sum over the data samples obtained from a training set. Indeed, the knowledge of the input distribution $\phi$ is indeed convenient, especially for the analysis. However, for the design it is not required. This is why the proposed GO quantization algorithm is applied to the problem of data clustering, in which only a database is available.} \textcolor{black}{The case of a time-varying input distribution is not addressed here and would require to design an adaptive quantizer, which is left as a relevant extension of the present work.} Also note that the set $\mathcal{X}$ and the function $\chi(g)$ are assumed to integrate the possible constraints on the decision $x$. At last, note that when the optimal decision function (ODF) $\chi(g)$ is not available, other decision functions that are suboptimal but easier to implement may be considered; this situation will be studied in the numerical analysis.  

In what follows, the main focus is on the regime of large $M$, which is called the high resolution regime. This regime is not only very useful to conduct the analysis and make interpretations but also to provide \textcolor{black}{neat approximants or expressions. These expressions are both exploited to obtain useful insights for the design of general quantizers and used in the proposed quantization algorithm. As it will be seen in the numerical performance analysis, the proposed algorithm performs remarkably well in the low resolution regime.} Note that the direct minimization of the general form of the OL is an NP-hard problem since it is a mathematical generalization of the conventional quantization problem (see e.g., \cite{Garey-TIT-1982,Hanna-JSAIT-2020}). Therefore, using approximants and suboptimal procedures is a classical approach in the area of quantization especially for vector quantization.




\section{Scalar GOQ in the high resolution regime}
\label{sec:scalar_approximation}

In this section we assume that both the decision to be taken and the parameter to be quantized are scalar that is, $d = p = 1$. For a wireless communication, this would occur for instance when a receiver has to report a scalar channel quality indicator (such as the SINR\textcolor{black}{, the carrier/interference ratio, or the received signal power}) to a transmitter and the transmitter tunes in turn its transmit power. Similarly, a real-time pricing system \cite{mohsenian-2010} in which an electrical power consumer reports its time-varying satisfaction parameter to an aggregator who chooses the price dynamically corresponds to the scalar case. Additionally, many systems, for complexity reasons, implement a set of independent scalar quantizers instead of a vector one. This is the case for example for some image compression standards such as JPEG or for MIMO communications with quantized CSI feedback \cite{Xu-TSP-2010,Makki-TCOMM-2013,Makki-TCOMM-2015}.
 \textcolor{black}{In the general case, finding a quantizer amounts to finding both the regions $\mathcal{G}_1,...,\mathcal{G}_M$ (which are just intervals in the scalar case) and the representatives $z_1,...,z_M$. However, the calculation of regions and representatives can be simplified in the HR regime. One could use probabilistic density function to represent the density of quantization points, which allows us to approximate summations by integrals.} To be precise, we assume the HR regime in the following sense \cite{Gray_TIT_1998}. For any point $g$, let us introduce the quantization step $  \Delta(g) = \min_{1\leq m \leq M} | g- z_m |  $. Then, let us introduce the (interval/representative) density function $ \rho(g) $ which is defined as follows:
\begin{equation}
    \rho(g)  = \lim_{M \rightarrow +\infty} \frac{1}{M \Delta(g)}. 
    \label{eq:definition_representative_density}
\end{equation}

\textcolor{black}{\subsection{Optimal quantization interval density function}}
By construction, the number of quantization intervals or representatives in any interval $[a,b]$ can be approximated by $M\displaystyle\int_a^b\rho(g)\mathrm{d}g$. Therefore, the problem of finding a GOQ in the HR regime amounts to finding the density function that minimizes the OL that we will denote, with a small abuse of notation but for simplicity by $L(\rho; f)$. Remarkably, the expression of the optimal density in the HR regime can be obtained, at least by assuming the goal and decision functions to be sufficiently regular or smooth. This is the purpose of the next proposition.

\begin{proposition}\label{prop:optimal-density} Let $f$ be a fixed goal function. Assume $f$ $\kappa$ times differentiable and $\chi$ differentiable with
\begin{equation}
    \kappa = \min  \left \{i \in \mathbb{N} : \left.\forall g,\,\,\frac{\partial^{i}f(x;g)}{\partial x^{i}} \right|_{x=\chi(g)}\neq 0 \,\,\mathrm{a.s.}  \right \}.
    \label{eq:k_definition}
\end{equation}
In the HR regime the OL $L(\rho; f)$ is minimized by using the following quantization interval/representative density function:

\begin{equation}
\rho^{\star}(g)=  C \left[\left(\frac{\mathrm{d}\chi(g)}{\mathrm{d} g} \right)^\kappa\frac{\partial^{\kappa}f(\chi\left(g\right);g)}{\partial x^{\kappa}}\phi(g) \right]^{\frac{1}{\kappa+1}}
\label{eq:lambda_op_general}
\end{equation}
where  \textcolor{black}{$\frac{1}{C} =  \displaystyle\int_\mathcal{G} \left [\left(\frac{\mathrm{d}\chi(t)}{\mathrm{d} t}\right)^\kappa\frac{\partial^{\kappa}f(\chi\left(t\right);t)}{\partial x^{\kappa}}\phi(t)\right ]^{\frac{1}{\kappa+1}}\mathrm{d}t $}.
\end{proposition}

\begin{proof}
See Appendix A.
\end{proof}
Although the optimal density is derived in the special case of scalar quantities and the HR regime, the corresponding result is insightful both for the analysis and the design. The conventional result when distortion minimization is pursued is that the optimal density $\rho^{\star}$ is proportional to $\phi^{\frac{1}{3}} \left(g \right)$. In practice this means allocating more quantization bits to more likely realizations of $g$. Under the GOQ \textcolor{black}{approach}, this conclusion is seen to be questioned. Indeed, the best density is seen to result from a combined effect of the parameter density $\phi$, the variation speed of $f$ w.r.t. the decision $x$ (that is, the sensitivity of the goal regarding the decision), and the smoothness of the decision function $\chi$ w.r.t. the parameter to be quantized. As a consequence all these three factors need to be acccounted for in practice to design a good GOQ and allocate quantization bits in particular. Let us illustrate this with a simple example that is relevant to the problem of energy-efficient wireless transmit power control.

\textit{Example.} Consider the following energy-efficiency (EE)  performance metric $f(x;g) = -\frac{\exp \left(-\frac{c}{xg} \right)}{x^{\eta}}$ with $c > 0$ and $\eta \geq 2$. Here $x$ represents the transmit power and $g$ the channel gain \cite{vero-TSP-2011} . Assume the channel gain $g$ is exponentially distributed that is, $\phi \left( g\right) = \frac{1}{\overline{g}} \exp \left(-\frac{g}{\overline{g}} \right)$ with $\mathbb{E}(g) = \overline{g} > 0$. One obtains that $\kappa=2$, $\chi(g) = \frac{c}{\eta g}$ and
\begin{equation}
     \rho^{\star}(g)=  C \left[  \frac{\eta^{\eta+1}}{c^\eta e^\eta}g^{\eta-2} \phi \left( g \right)   \right]^{\frac{1}{3}}.
\end{equation}
For instance, for $\eta=3$, it is easy to check that the quantization interval density $\rho^{\star}$ is increasing for $0 \leq g \leq \overline{g}$  then decreasing for $ g \geq \overline{g}$. This result thus markedly differs from the conventional distortion-based approach. Indeed, under the latter approach, one would allocate more quantization bits to small values of the channel gain (since $\phi$ is strictly decreasing). Under the GOQ approach, most of the allocation bits should be allocated for values around the mean value of $g$. 

In this section, we have been searching for the best scalar GOQ for a given goal function $f$. Now, we would like to provide some elements about the relationship between the nature of $f$ and the quantization performance. For example, it is known that compressing a signal for which its energy is concentrated at small frequencies is generally an easy task. Similarly, here, we would like to know more about the connection between the regularity properties of the goal function and the level of difficulty to quantize its parameters. Since, this relevant issue constitutes a challenging mathematical problem, we only provide some preliminary results to explore this promising direction. For this purpose, we assume the chosen quantizer to be given by the optimal HR quantizer given by $\rho^{\star}$ and study the impact of $f$ on $L(\rho^{\star}; f)$. To be rigorous and clearly indicate the dependency of $\rho^{\star}$ regarding $f$, we will use the notation $ \rho^{\star}_f $.

\textcolor{black}{\subsection{About choosing the scaling factor $\alpha_f$}}
So far, since $f$ was fixed, the scaling factor $\alpha_f$ in the definition of the OL $L$ was not relevant. But when it comes to minimizing $L(\rho_f^{\star}; f)$ w.r.t $f$, this factor plays an important role. Indeed, if one wants to compare the hardness to compress of two functions, the retained performance criterion has to possess some invariance properties. In particular, it should be invariant to affine transformations. The OL has not this property regarding $f$ since a function of the form $F = A f +B$ (with $A>0$) would produce a large OL when $A$ is large even if the OL obtained for the original $f$ is small. Hence the need for normalizing the OL properly and thus the presence of $\alpha_f$. Here, we consider two choices for $\alpha_f$, which amounts to considering two different reference case for the performance comparison. The first reference case is uniform quantization. For this case, the normalizing factor is denoted by $\alpha_f^{\mathrm{UQ}}$ and chosen to be the reciprocal of the OL obtained when using a HR uniform quantizer (UQ). It expresses as:
\begin{equation}
    \frac{1}{\alpha_f^{\mathrm{UQ}}} =  \displaystyle\int_{\mathcal{G}} C_g^{-\kappa}\left(\frac{\mathrm{d}\chi(g)}{\mathrm{d} g}\right)^\kappa\frac{\partial^{\kappa}f(\chi(g);g)}{\partial x^{\kappa}}\phi(g)\mathrm{d}g
\end{equation}
where $\displaystyle \int_{\mathcal{G}}C_g\mathrm{d}g=1$. This case allows one to quantify the potential gain from using a GOQ instead of a standard quantizer which is independent of the goal function. The second reference case we consider corresponds to the situation where the DM entity takes a constant decision (CD) \textcolor{black}{independently of} the value of $g$. This would correspond to the situation where no instantaneous information about $g$ is available and only statistics can be exploited. Although this reference case is not necessarily the right benchmark for a given application it is still of interest for extracting useful insights because, this time, it is not about comparing two quantizers but more about measuring the intrinsic difficulty to compress a given function. By defining $\overline{x}$ the 
chosen constant decision as $\widebar{x} \in\underset{x \in \mathcal{X}}{\arg\min}\  \mathbb{E}_g \left[ f \left(x;g \right) - f \left(\chi(g); g\right) \right]$, the corresponding normalizing factor is denoted by $\alpha_f^{\mathrm{CD}}$ and expresses as:
\begin{equation}
    \frac{1}{\alpha_f^{\mathrm{CD}}}=  \frac{1}{(2M)^{\kappa} \kappa! (\kappa+1)} \displaystyle \int_{g \in \mathcal{G}} \left[ f \left(\widebar{x};g \right) - f \left(\chi \left(g\right);g \right) \right] \phi \left( g\right)\mathrm{d}g   
\end{equation}
where $\overline{x}$ is the chosen constant decision. The above quantity represents the OL obtained when using the best CD multiplied par a term in $\kappa$ which comes from the HR approximation (see App. A for more details).

\textcolor{black}{\subsection{On the impact of the goal function on the OL}}
Equipped with these two versions of the (normalized) OL, comparing different goal functions becomes a well posed problem. For this purpose, we have selected several functions \cite{vero-TSP-2011,Meshkati-JSAC-2007,Berri-EUSIPCO-2016} that frequently appear in wireless resource allocation problems. For the selected functions, all quantities at hand can be expressed analytically and the integral associated with the OL can be computed. The obtained results appear in  \textcolor{black}{Table \ref{tab:scalr_cost_function_comparison}}. \textcolor{black}{With the parameter space  taken to be the interval $[0.1,10]$, the table assumes two different choices for the p.d.f. $\phi$, the uniform distribution and a {truncated} exponential distribution namely, $\phi(g)=\frac{\exp(-g)}{\displaystyle \int_{0.1}^{10}\exp(-x)\mathrm{d}x}$}. The two columns providing the value of the OL allows one to establish some hierarchy between the selected functions. The obtained results suggest that logarithm-type goal functions provide a relatively small OL. These types of function would be qualified as easy to compress, which means for example that a rough description of the parameter is sufficient to take a good decision. Quantizing finely the parameter would lead to a waste of resources. This interpretation which is based on the HR analysis will be confirmed by simulations performed in arbitrary regimes. In a wireless system, this would e.g., mean that transmission rate-type performance metrics are not very sensitive to quantization noise and therefore a coarse feedback on CSI is suited to the goal. The table shows a different behavior for exponential-type functions, which are typically used to model energy-efficiency in wireless systems. These types of function require a more precise description of the function parameters (e.g., the CSI). Implementing the GOQ approach for such functions is seen to still provide a quite significant gain in terms of OL when compared to uniform quantization. We see that the HR analysis of the scalar quantization case provides useful insights that could be both used for an ad hoc design of a goal-oriented quantizer and deepened by considering more complex performance metrics. 


\begin{table}[tbp]
\centering
\begin{tabular}{|p{1.8cm}|p{1.05cm}|p{1.28cm}|p{1.25cm}|p{1.25cm}|}
\hline
\textbf{{Goal function $f(x;g)$}} & {$\textbf{p.d.f.}$~$\phi\left(g \right)$} &{$\textbf{ODF}$~$\chi\left(g \right)$}  & \textbf{OL} ($\alpha_f=  \alpha_f^{\mathrm{UQ}}$) & \textbf{OL} ($\alpha_f=  \alpha_f^{\mathrm{CD}}$) \\
\hline
{$\log \left(1+10gx \right)-x$} & {{uniform}}& {$[1-\frac{1}{10g}]^+$}   & {$0.00399$}&  $0.0488$\\
\hline
{$\frac{\exp(-\frac{1}{gx})}{x}$} & {{uniform}} & {$\frac{1}{g}$}  & {$0.648$} &  $6.5943$\\
\hline
{$\frac{(1-\exp(-{gx}))^{10}}{x}$} &{{uniform}} & {$\frac{3.6150}{g}$}  & {$0.648$} &  $19.4565$\\
\hline
{$(x-g)^2$} &uniform & {$g$}   & {$1$}&  $24$ \\
\hline
{$\log \left(1+10gx \right)-x$} & {{exp}}& {$[1-\frac{1}{10g}]^+$}   & {$0.0019$} & 0.4859\\
\hline
{$\frac{\exp(-\frac{1}{gx})}{x}$} & {{exp}} & {$\frac{1}{g}$}  & {$0.083$}& 18.75\\
\hline
{$\frac{(1-\exp(-{gx}))^{10}}{x}$} &{{exp}} & {$\frac{3.6150}{g}$}  & {$0.083$}& 61.12\\
\hline
{$(x-g)^2$} &exp & {$g$}   & {$0.24$} & 48.50\\
\hline
\end{tabular}

\caption{Comparison of different goal functions \label{tab:scalr_cost_function_comparison}}
\end{table}

\section{Vector GOQ: High resolution analysis and proposed quantization algorithm}
\label{sec:vector_approximation}

\subsection{High resolution analysis}

As motivated in Sec.~\ref{sec:scalar_approximation}, for some applications vector quantization is not used for reasons such as computational complexity. This is the case for instance for MIMO systems where the transfer channel matrix entries are  quantized by a set of scalar quantizers. But, for optimality reasons or because of the definition of the quantization problem, vector quantization may be necessary. For instance, it is of high practical interest to be able to cluster series of the non-flexible electrical power consumption over one day for example \cite{beaude-tsg}\cite{clustering}\cite{Zhang-AE-2021}, which leads to a sample dimension of $p=48$ when the power signal is sampled every $30$ minutes. By construction, this clustering problem is similar to a vector quantization problem for which one wants to create a certain number ($M$ with our notation) of data subsets. For this specific problem one may want to fix $M$ to a small number, say $M=4$, and distinguish between $4$ consumption behaviors. 

For the scalar case, it has been seen that the HR regime allows to determine the best goal-oriented quantizer, which is fully characterized by the density function $\rho^{\star}$ (see (\ref{eq:lambda_op_general})). However, in the vector case, even under the HR assumption, the problem remains challenging in general. This is one of the reasons why we resort to approximations. The full analytical characterization of the corresponding approximations is left as a relevant extension of the present work. The goal in this paper is threefold: to show how these approximations can be used for the quantizer design; to support the choices made by simulations performed with a low and moderate number of quantization bits; to focus on the potential gains that can be brought by the GOQ approach. One the main results of this section consists in providing an exploitable approximation of the OL in the vector case. This approximation will be directly exploited further in this section for the quantizer design part. The result is stated \textcolor{black}{through} the following proposition.

\begin{proposition} 
\label{prop:upper_lower_bound}
Assume $d\geq 1$, $p \geq 1$, and $\kappa=2$. Assume $f$ and $\chi$ twice differentiable. Denote by $\mathbf{H}_f(x;g)$ the Hessian matrix of $f$ and denote by $\mathbf{J}_{\chi}(g)$ the Jacobian matrix of $f$ evaluated for an optimal decision $\chi(g)$. In the regime of large $M$, the optimality loss function $L(Q;f)$   \textcolor{black}{defined as} in (\ref{eq:def-OL}) can be approximated as follows:

{\footnotesize\begin{equation}
 L(Q;f) = \underbrace{\alpha_f \sum_{m=1}^M \int_{\mathcal{G}_m} (g - z_m)^{\mathrm{T}} 
\mathbf{A}_{f,\chi}(g) (g - z_m) \phi(g) \mathrm{d}{g}}_{\widehat{L}_M(Q;f)} + \textcolor{black}{o(M^{-\frac{2}{p}})}
\end{equation}}
where 
$\mathbf{A}_{f,\chi}(g) =\mathbf{J}_{\chi}^{\mathrm{T}}(g) \mathbf{H}_f(\chi(g);g) \mathbf{J}_{\chi}(g)$. Additionally, by assuming the Gersho hypothesis \cite{Gersho_TIT_1979} (see App. B), the above first order HR equivalent of $L$ can be bounded as \textcolor{black}{$L_M^{\min}((Q;f) \leq \widehat{L}_M(Q;f) \leq L_M^{\max}(Q;f) $} with

\begin{equation}
 \textcolor{black}{L_M^{\min}(Q;f)} =
\frac{p \mu_p}{2} M^{-\frac{2}{p}}\left(\displaystyle\int_{\mathcal{G}}\left(\lambda_{\min}(g;f) \phi({g})\right)^{\frac{p}{p+2}}\mathrm{d}g\right)^{\frac{p+2}{p}}
\label{eq:lower_bound_vec}
\end{equation}
\begin{equation}
\textcolor{black}{L_M^{\max}(Q;f)}=
\frac{p \mu_p}{2} M^{-\frac{2}{p}}\left(\displaystyle\int_{\mathcal{G}}\left(\lambda_{\max}(g;f) \phi({g})\right)^{\frac{p}{p+2}}\mathrm{d}g\right)^{\frac{p+2}{p}}
\label{eq:upper_bound_vec}
\end{equation}

where: $\lambda_{\min}(g;f)$ (resp. $\lambda_{\max}(g;f)$) is the smallest (resp. largest) \textcolor{black}{eigenvalue} of $\mathbf{A}_{f,\chi}(g)$ and \textcolor{black}{$\mu_{p}$ is the least normalized moment of inertia of the $p$-dimensional tessellating polytope $\mathbb{T}_p$ defined by
\begin{equation}
\mu_p=\min_{\mathbb{T}_p,z}\frac{1}{p}\frac{1}{\mathrm{vol}(\mathbb{T}_p)^{1+2/p}}\displaystyle\int_{\mathbb{T}_p}\|{g}-{z}\|^2\mathrm{d}{g}.
\end{equation}}

\end{proposition}
\begin{proof}
See Appendix B.
\end{proof}
The first-order equivalent in Prop. \ref{prop:upper_lower_bound} is seen to depend on the matrix $\mathbf{A}_{f,\chi}(g)$. This matrix corresponds to the vector generalization of the product $\left(\frac{\mathrm{d}\chi(g)}{\mathrm{d} g} \right)^2\frac{\partial^{2}f(\chi\left(g\right);g)}{\partial x^{2}}$ that appears in the scalar case and shows how the OL is related to the regularity properties of the goal function $f$. For the conventional quantization approach ($f(x;g) = \|x-g \|^2$), one has merely that $\mathbf{A}_{f,\chi}(g) = \mb{I}$. Therefore in the HR regime, the structure of the equivalent shows that considering a general goal function $f$ amounts to introducing an appropriate weighting matrix in the original distortion function. This matrix will be precisely used to derive an algorithm to compute a good vector GO quantizer that is tailored to the goal function.  

The derived lower and upper bounds can be used both for characterizing the performance of a GOQ and for the quantizer design, which is explained at the end of this section. The bounds are tight in special cases such as when $p=1$ (in which case $\mu_p = \frac{1}{12}$) and when $f(x;g) = \|x-g \|^2$ (with no restrictions on the dimensions $d$ and $p$. Generally speaking, the gap between the two bounds is observed to be small when $p$ is less or much less than $d$.  Now if $p\geq d$, it can be seen that $\lambda_{\min}(g;f) = 0$ since the matrix $\mathbf{A}_{f,\chi} \left( g\right)$ is not full rank. As a consequence, the lower bound derived in (\ref{eq:lower_bound_vec}) is not tight anymore. Hence, it is necessary to derive a tighter lower bound in this scenario. To this end, one can treat $\mathbf{J}_{\chi}(g) e_m$, with $e_m=\frac{{g}-{z}_m }{\|{g}-{z}_m\|}$, as a vector and thus ${e}_m^{\mathrm{T}}\mathbf{A}_{f,{\chi}} \left({g}\right) {e}_m$ is minimized if and only if $\mathbf{J}_{{\chi}}({g}){e}_m$ is aligned with the eigenvector associated with the smallest eigenvalue of $\mathbf{H}_f({\chi} \left( {g}\right);{g})$. By denoting $\nu_{\min}({g};f)$ the smallest eigenvalue of \textcolor{black}{$\mathbf{H}_f({\chi} \left( {g}\right); {g})$}, the term ${e}_m^{\mathrm{T}}\mathbf{A}_{f,{\chi}} \left( {g}\right) {e}_m$ can be lower bounded by $\nu_{\min}({g};f)\mathsf{a}(\mathbf{J}_{{\chi}}({g}))$, where $\mathsf{a}(\text{J}_{{\chi}}({g}))$ is the scalar factor between $\mathbf{J}_{{\chi}}({g}){e}_m$ and the smallest eigenvector of $\mathbf{H}_f({\chi} \left({g}\right);{g})$. By replacing $\lambda_{\min}(g;f)$ with $\nu_{\min}(g;f)\mathsf{a}(\mathbf{J}_{{\chi}}({g}))$, a new lower bound can be derived for the case where $p\geq d$. The proposed refinement procedure can also be used for the upper bound on the OL but note that the upper bound is mainly dependent on $p$ and is much less dependent on the dimensionality $d$, which makes the corresponding refinement generally less useful.

\subsection{Proposed quantization algorithm}
As mentioned in the last subsection, the bounds provided by Prop.~\ref{prop:upper_lower_bound} can be used to characterize the performance of a quantizer and study, at least numerically, the impact of the nature of $f$ on the OL. In the present subsection, the main objective is to exploit the HR equivalent of Prop.~\ref{prop:upper_lower_bound} to design a practical quantization algorithm. Considering the fact that the optimal decision function may produce solution at the boundary of the decision set and that only sub-optimal decision function may be available in real systems, we relax here the optimality first order condition  $\frac{\partial f(x;g)}{\partial x }|_{x=\chi(g)}=0$. Therefore, the optimality loss can be written for algorithmic purposes in a more general form:
\begin{equation}
\begin{split}
&L(Q;f) \\
= & \sum_{m=1}^M \int_{\mathcal{G}_m}  \left[\left(\frac{\partial f(x;g)}{\partial x}\left|_{x=\chi(g)}\right.\right)^{\mathrm{T}} \right. \left(\chi(z_m)-\chi(g)\right)\\ 
&+ \left.\frac{1}{2}\left(\chi(z_m)-\chi(g)\right)^{\mathrm{T}} 
\mathbf{H}_{f,\chi}(g) \left(\chi(z_m)-\chi(g)\right) \right]\phi(g) \mathrm{d}g\\
&+o\left(\|\chi(z_m)-\chi(g)\|^2\right) 
\end{split}
\end{equation}
where $\left(\mathbf{H}_{f,\chi}(g) \right)_{i,j}=\frac{\partial^2 f(x;g)}{\partial x_i \partial x_j}|_{x=\chi(g)}$ for $1 \leq i,j \leq p$. By using the Taylor expansion, we have that:
\begin{equation}
\begin{split}
     &\chi(z_m)-\chi(g)\\= &\mathbf{J}_{\chi}(g)(z_m-g)+\begin{bmatrix}
(z_m-g)^{\mathrm{T}}\mathbf{H}_{\chi_1}(g)(z_m-g)\\
(z_m-g)^{\mathrm{T}}\mathbf{H}_{\chi_2}(g)(z_m-g)\\
\dots\\
(z_m-g)^{\mathrm{T}}\mathbf{H}_{\chi_{d}}(g)(z_m-g)\\
\end{bmatrix}\\
+&o\left(\|z_m-g\|^2\right)
\end{split}
\end{equation}
where $ \left( \mathbf{H}_{\chi_i} \left(g \right) \right)_{\ell,k}=  \frac{\partial^2\chi_i(g)}{\partial g_\ell \partial g_k} $  for $1\leq l,k \leq p$ and $1\leq i \leq d$, $\chi(g)=[\chi_1(g),\dots,\chi_d(g)]^{\mathrm{T}}$. Plugging this expression in the expression of $L(Q;f)$, the optimality loss can be re-expressed as
\begin{equation}
\begin{split}
&L(Q;f)\\=&\frac{1}{2} \sum_{m=1}^M \int_{\mathcal{G}_m} (g - z_m)^{\mathrm{T}} \mathbf{B}_{f,\chi}(g) (g - z_m) \phi(g) \mathrm{d}{g}\\+&\frac{1}{2} \sum_{m=1}^M \int_{\mathcal{G}_m} (g - z_m)^{\mathrm{T}} \mathbf{A}_{f,\chi}(g) (g - z_m) \phi(g) \mathrm{d}{g} + o\left(M^{\frac{-2}{p}}\right)
\end{split}
\end{equation}
where $\mathbf{B}_{f,\chi}(g)=\sum_{i=1}^{d}\bigtriangledown f_i(g)\mathbf{H}_{\chi_i}(g)$ with\[\frac{\partial f(x;g)}{\partial x}|_{x=\chi(g)}=\left(\bigtriangledown f_1(g),\bigtriangledown f_2(g),\dots,\bigtriangledown f_d(g)\right)\] and $\mathbf{A}_{f,\chi}(g) =\mathbf{J}_{\chi}^{\mathrm{T}}(g) \mathbf{H}_f(x;g) \mathbf{J}_{\chi}(g)$.

By using this new expression of the OL, one exhibits a natural structure for applying an alternating optimization algorithm and thus to minimize  $\widetilde{L}=\sum_{m=1}^M \displaystyle\int_{\mathcal{G}_m} (g - z_m)^{\mathrm{T}} \left(\mathbf{B}_{f,\chi}(g)+\mathbf{A}_{f,\chi}(g)\right) (g - z_m) \phi(g) \mathrm{d}{g}$ as follows:
\begin{itemize}
    \item {\textit{Representative updating step}: To minimize $\tilde{L}$ with \textbf{fixed regions}, the problem boils down to find the representative $z_m$ such that $\displaystyle \int_{\mathcal{G}_m} (g - z_m)^{\mathrm{T}} \left(\mathbf{B}_{f,\chi}(g)+\mathbf{A}_{f,\chi}(g)\right) (g - z_m) \phi(g) \mathrm{d}{g}$ can be minimized. 
    One can apply a gradient descent technique to achieve that where the gradient can be easily found:
    \begin{equation}
        \frac{\partial \tilde{L}}{\partial z_m} = 2 \displaystyle \int_{\mathcal{G}_m} \mathbf{E}_{f,\chi}\left(g \right) \left(g-z_m \right) \phi \left( g\right) \mathrm{d}g
    \end{equation}
    where $\mathbf{E}_{f,\chi}\left(g \right) = \mathbf{B}_{f,\chi}\left(g \right) + \mathbf{A}_{f,\chi}\left(g \right)$.
    }
   
    \item{\textit{Region updating step}: For given representatives, the region can be computed as:\[\begin{split}\mathcal{G}_m=&\ \left\{ g|(g - z_m)^{\mathrm{T}} \mathbf{E}_{f,\chi}(g) (g - z_m) \right.\\& \left.\leq (g - z_{m'})^{\mathrm{T}} \mathbf{E}_{f,\chi}(g) (g - z_{m'})\right\}\end{split}\] where $m'\neq m$.}
\end{itemize}
The approximate individual optimality loss is thus defined by $\widetilde{\ell}_f \left(g,z \right)$ of the parameter $g$ w.r.t. a representative $z$ as:
\begin{equation}
\widetilde{\ell}_f \left(g,z \right) \triangleq \left(g - z \right )^{\mathrm{T}} \mathbf{E}_{f,\chi}(g) \left(g - z\right).
\end{equation}
our goal-oriented quantization algorithm is summarized in pseudo-code form through \textcolor{black}{algorithm} \ref{alg:goq_gradient}. \textcolor{black}{The proposed algorithm can be applied to the scalar case. In the latter case, the matrix $\mathbf{A}_{f,\chi}\left(g \right)$ becomes $\left(\frac{\mathrm{d}\chi(g)}{\mathrm{d} g} \right)^2 \frac{\partial^{2}f(\chi\left(g\right);g)}{\partial x^{2}}$ which corresponds to the term appearing in Equation \ref{eq:lambda_op_general} with $\kappa = 2$. And we have that $\mathbf{B}_{f,\chi}\left(g \right) = 0 $. The reason for this is that either the first-order optimality condition holds or the lower and upper bounds of the quantization interval are fixed points.}
\begin{algorithm}[tbp]
{\bf{Inputs:}} goal function $f\left(x;g \right)$, $\chi \left(g \right)$, error tolerance $\varepsilon$, \textcolor{black}{number of cells $M$} and number of iterations
$T$;\\
{\bf{Inputs:}}  $\mathcal{Z}^{\left(0\right)}=\left\{ {z}_{1}^{\left(0\right)},\dots,{z}_{M}^{\left(0\right)}\right\}$;\\
{\bf{Inputs:}}  $\mathcal{G}^{\left(0\right)}=\left \{ \mathcal{G}_{1}^{\left(0\right)},\dots,\mathcal{G}_{M}^{\left(0\right)}\right\};$\\

\For{$t = 1$ \bf{to} $T$} {
\For{$m = 1$ \bf{to} $M$}{
 Update $\mathcal{G}^{\left(t\right)}_m $ by $\left \{ g \left |\widetilde{\ell}_f \left(g,z^{(t-1)}_m \right) \leq \widetilde{\ell}_f \left(g,z^{(t-1)}_{m'} \right),  \forall m' \neq m \right. \right\}$;\\
 Update ${z}^{\left(t\right)}_m$ by 
 ${z}^{\left(t\right)}_m = {z}^{\left(t-1\right)}_m - r_t \frac{\partial \widetilde{L} \left(\mathcal{Z}^{\left(t-1\right)} \right)} {\partial z^{\left(t-1\right)}_m}$ with the step size $r_t > 0$ s.t. ${z}^{\left(t\right)}_m \in 
 \mathcal{G}$;\\ 
}
\If{$\sum_{m=1}^{M}\left\Vert {z}_{m}^{\left(t\right)}-{z}_{m}^{\left(t-1\right)}\right\Vert ^{2}<\varepsilon$}{
\bf{Break};
}
}
{\bf{Outputs:}} \textcolor{black}{$\mathcal{Z}^{\star} = \mathcal{Z}^{\left(t\right)}$}  and $\mathcal{G}^{\star} = \mathcal{G}^{\left(t\right)}$; \\
\caption{ Goal-oriented Quantization Algorithm} \label{alg:goq_gradient}
\end{algorithm}



\section{Numerical performance analysis}
\label{sec:Numerical_Results}

In this section we both want to illustrate some analytical results derived in the preceding sections and also see, from purely numerical results, to what extent some insights obtained from the HR analysis hold in scenarios where main assumptions such as smoothness are relaxed. For this purpose, we consider four goal functions: an exponential-type goal function and a log-type goal function which are relevant for GO information quantization problems in wireless resource allocation problems; a quadratic-type goal function which is typically relevant for GOQ in controlled systems; an $\mathrm{L}_\mathrm{P}$ norm-type goal function which is relevant for GO data clustering/quantization in power systems. 

\textcolor{black}{\subsection{Impact of the goal function on the OL for wireless metrics}}

\begin{figure}[tbp]
\begin{centering}
\includegraphics[scale=0.4]{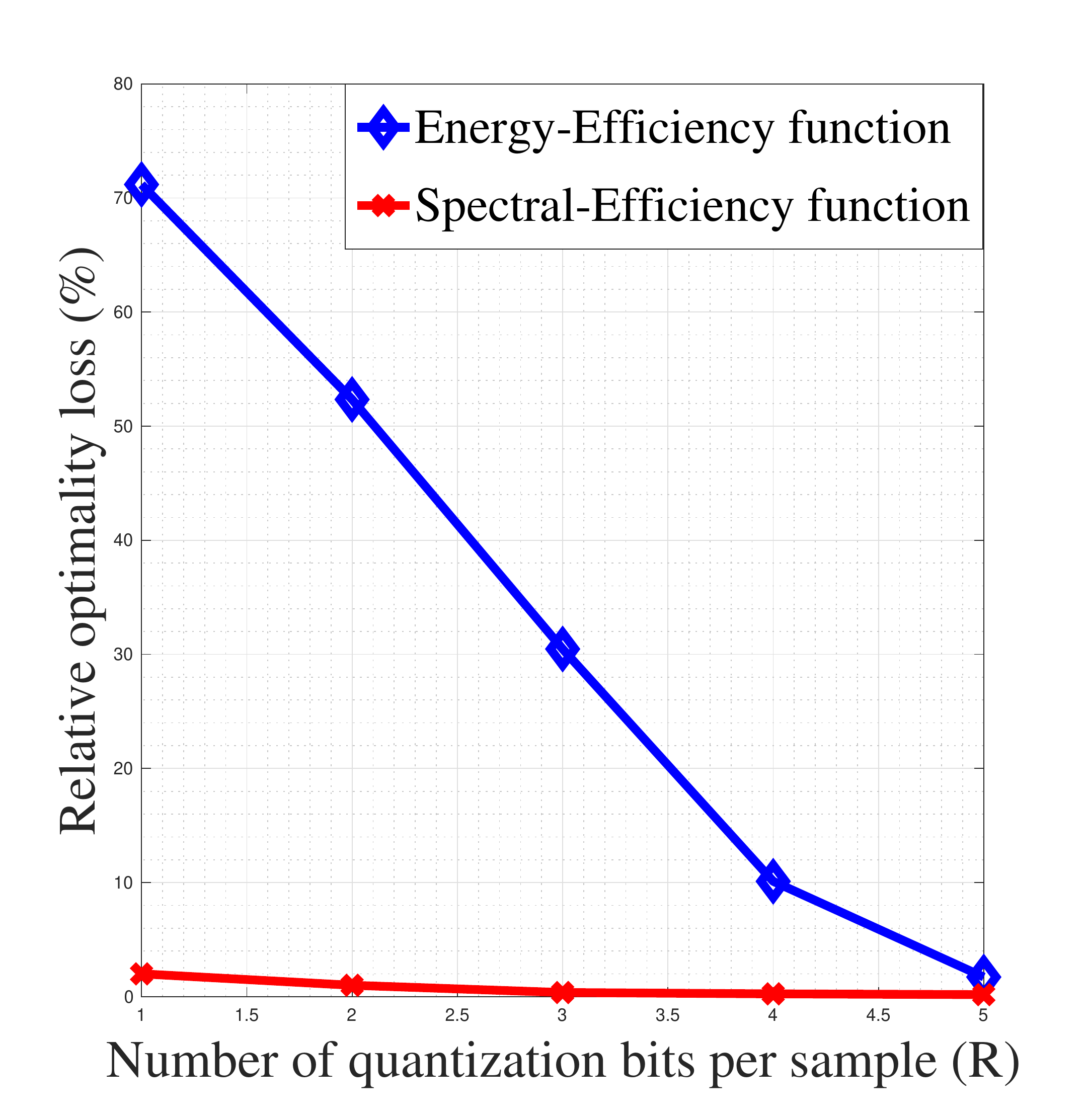}
\par\end{centering}

\caption{The figure shows the impact of the number of quantization bits on the decision-making quality (measured in terms of optimality loss) on two different well-used goal functions. Log-type SE functions appear to accommodate very well with very rough quantization for its parameters (CSI) which is the not the case for exp-type EE functions. This simulation is in accordance with the analytical results of Table I. 
\label{fig:ee_vs_sum_rate}
}
\end{figure}

Table \ref{tab:scalr_cost_function_comparison} provides analytical results for the scalar case in the HR regime. It suggests that for a given quantization scheme, log-type goal functions lead smaller values for the OL than exp-type goal functions. Let us consider the performance metric introduced by \cite{Meshkati-JSAC-2007} to measure the EE of a multiband communication: $f^{\text{EE}}\left(x;g \right) =- \frac{\sum_{i=1}^{S} \exp\left(-\frac{c}{x_i g_i} \right)}{\sum_{i=1}^{S} x_i}$ where $S$ is the number of bands, $c>0$, $x_i$ is the transmit power for band $s$, and $g_i$ the channel gain for band $s$. The log-type function is taken to be the classical spectral efficiency (SE) function $f^{\text{SE}} \left(x;g \right) = - \sum_{i=1}^{S} \log \left(1+ x_i\frac{g_i}{\sigma^2} \right)$. We impose that $x_i \geq 0 $ and  $\sum_{i=1}^{S} x_i \leq P_{\max}$. For $\frac{P_{\max}}{\sigma^2} = 5$, $c=1$, $S = 2$, and a uniform quantizer Fig.~\ref{fig:ee_vs_sum_rate} depicts the relative OL in percentage (relatively to the \textcolor{black}{ideal} case):
\begin{equation}
    \mathrm{Relative \ OL}(\%) = 100 \times \left(\frac{f(\chi(Q(g));g) -   f(\chi(g);g) }{f(\chi(g);g)} \right) 
\end{equation}
averaged over $10000$ independent Rayleigh fading realizations (with $\mathbb{E}(g)=1$) against the number of quantization bits per realization of $g$. We see for a given number of bits per sample, the OL for the SE function is much smaller than the SE function. We retrieve the hierarchy suggested by Table \ref{tab:scalr_cost_function_comparison}. This shows that the SE function can accommodate a rough quantization of the parameters (that is, the channel gains) without degrading significantly the DM process, which is to choose a good power allocation vector. Using a fine quantizer would lead to waste of resources for the SE function (here we see that a 1-bit quantizer yields an OL of about $2\%$, which illustrates well the importance of adapting the quantizer to the goal function. 

\textcolor{black}{\subsection{Performance gains obtained from tailoring the quantizer to the (control) goal}}

\begin{figure}[tbp]
\begin{centering}
\includegraphics[scale=0.44]{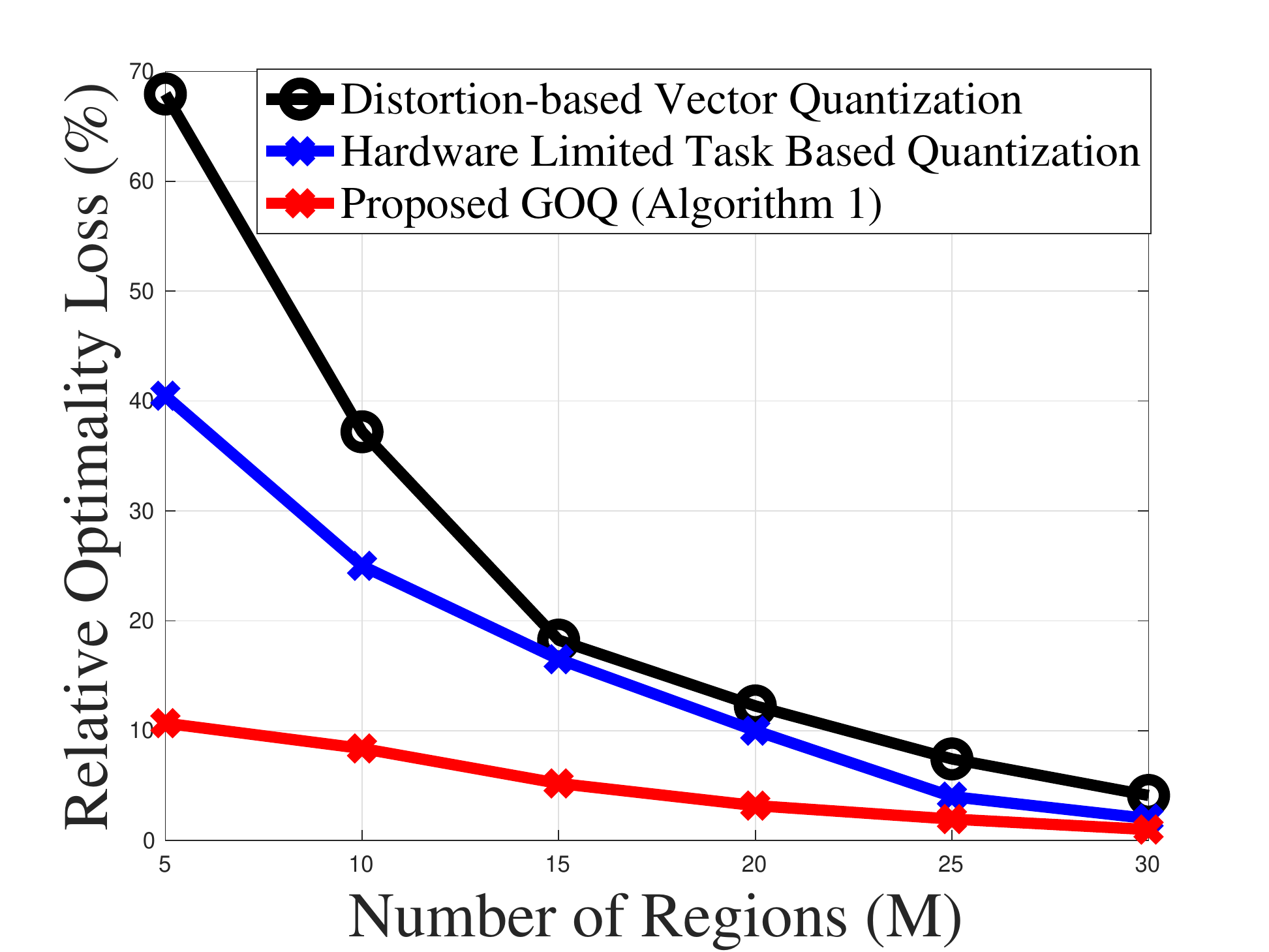}
\par\end{centering}
\caption{The goal function being a quadratic function, the figure shows the importance in terms of (decision) optimality loss of adapting the quantizer to the goal instead of using the conventional distortion-based quantization \textcolor{black}{approach}. \label{fig:ROL_quadratic}}
\end{figure}

\begin{figure}[tbp]
\begin{centering}
\includegraphics[scale=0.44]{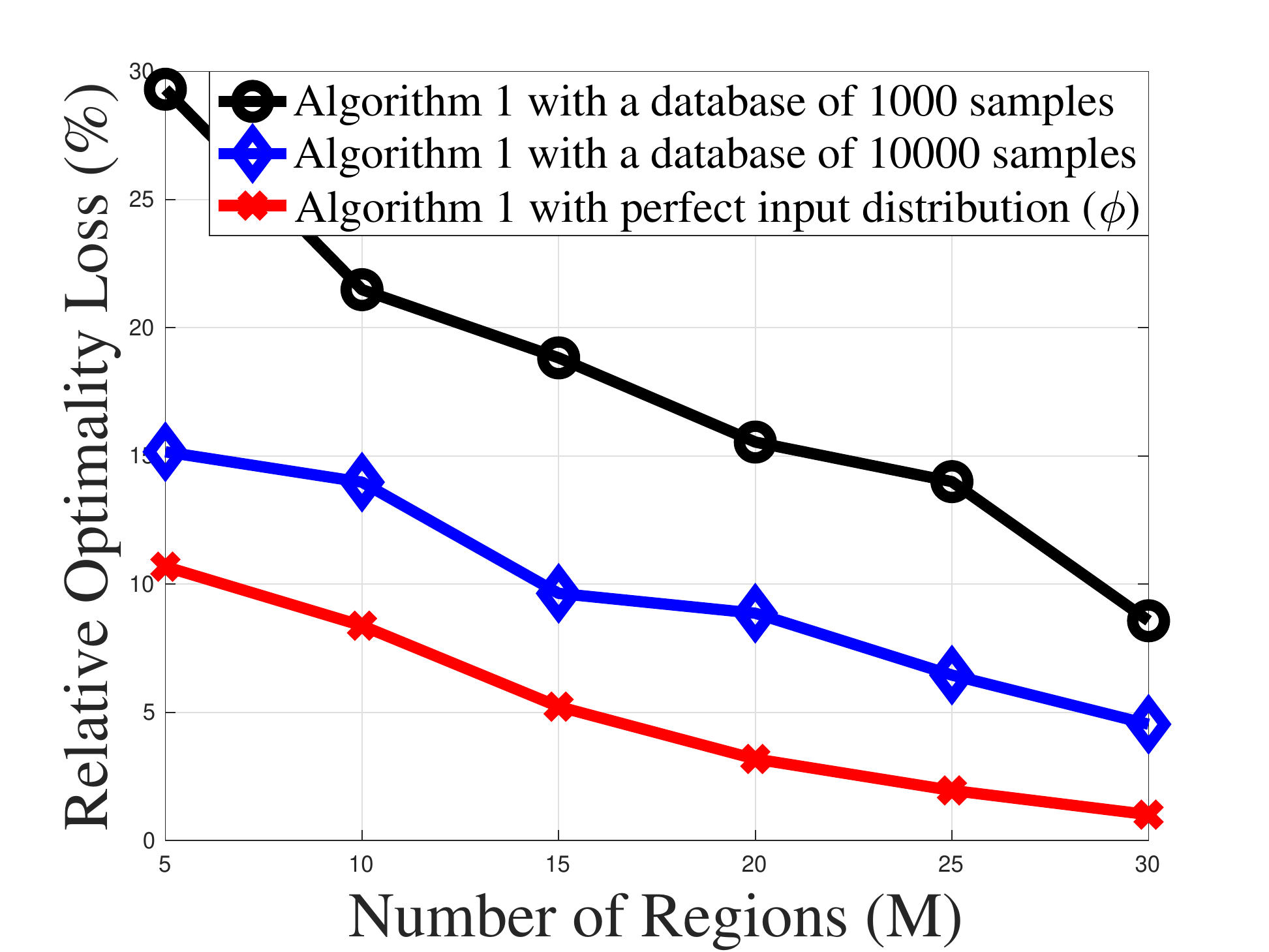}
\par\end{centering}
\caption{ \textcolor{black}{The figure assesses the performance loss due to not knowing the input distribution $\phi$ perfectly but rather with a low number of samples took from a database.} \label{fig:quadratic_data_oriented}}
\end{figure}

\begin{figure}[tbp]
     \centering
     \begin{subfigure}[b]{0.4\textwidth}
         \centering
         \includegraphics[width=\textwidth]{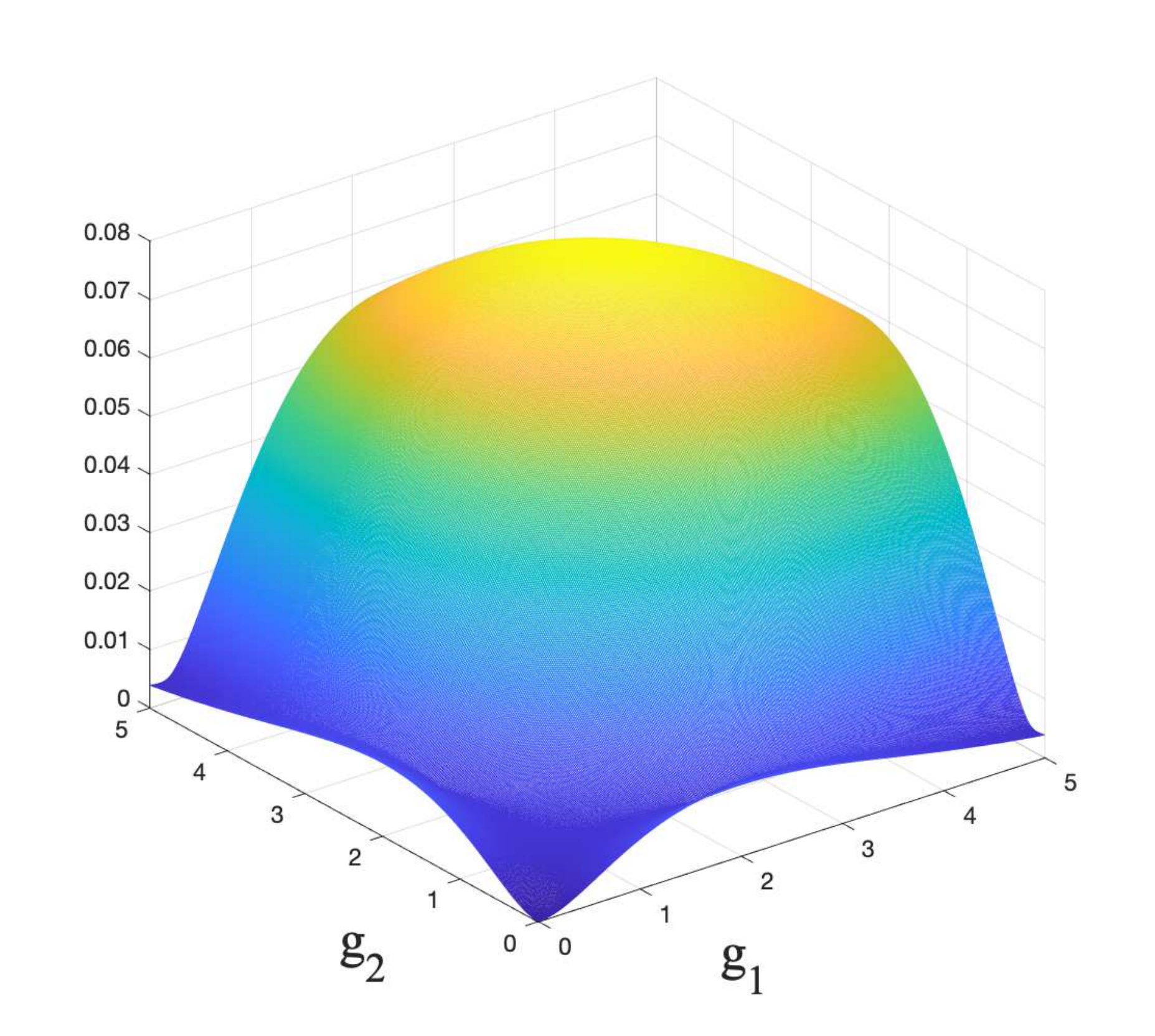}
         \caption{New density $\lambda_{\max} \left(g; f^\text{QUA} \right) \phi \left( g\right)$ of GOQ algorithm }
         \label{fig:pdf_goq}
     \end{subfigure}
     \hfill
     \begin{subfigure}[b]{0.4\textwidth}
         \centering
         \includegraphics[width=\textwidth]{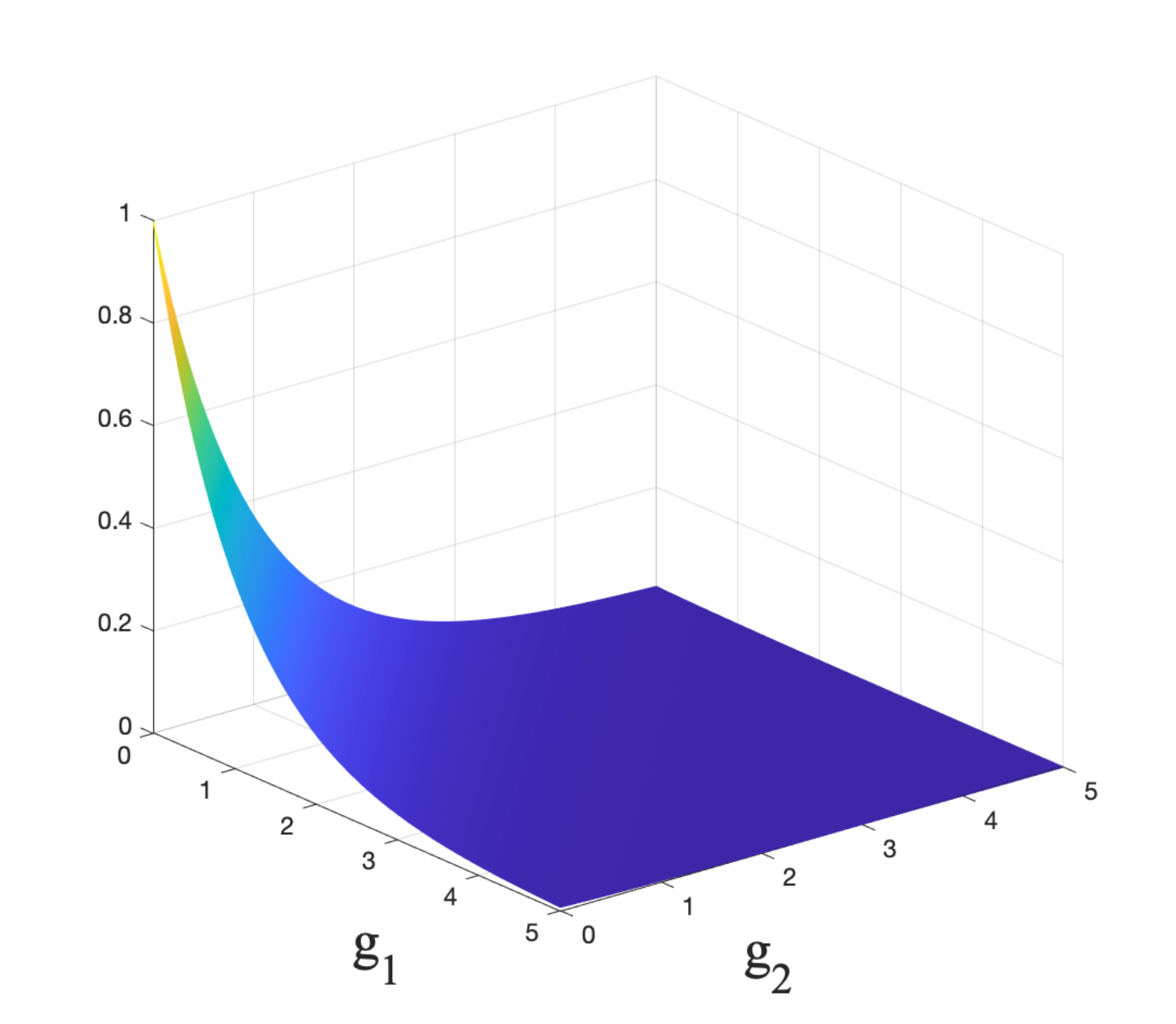}
         \caption{Original probability distribution $\phi \left(g \right)$}
         \label{fig:pdf_org}
     \end{subfigure}
        \caption{The figure shows the marked difference between the parameter probability distribution (bottom curve) and the probability distribution of interest that is relevant to the decision-making task (top curve). It implies in particular that quantization regions (and thus quantization bits) should be allocated in a very different way from the conventional way.}
        \label{fig:pdf_comparison}
\end{figure}

Now we assume $d = p = 2$ and consider the following quadratic function: 
\begin{equation}
    f^{\text{QUA}}(x;g) = \left(x_1 - h_1 (g) \right)^2 + \left({x}_2 - h_2 ({g}) \right)^2 + \left({x}_1 - {x}_2 \right)^2 \label{eq:cost_function_poly_2}
\end{equation} with $h_1({g})= 2{g}_1 {g}_2-\frac{1}{2} {g}_1^2 {g}_2^2 $ and $h_2({g}) = {g}_1^2 {g}_2^2 - {g}_1 {g}_2 $. Parameters are assumed to be i.i.d. and exponentially distributed, i.e., $\phi\left(g \right) = \exp\left(-g_1-g_2\right)$. One can check that  $\chi({g}) = [{g}_1 {g}_2, \frac{1}{2} {g}_1^2 {g}_2^2]^{\mathrm{T}}$. In Fig. \ref{fig:ROL_quadratic}, the relative OL in percentage (relatively to the \textcolor{black}{ideal} case) against the number of regions $M$ is represented for a conventional vector quantizer (namely, a distortion-based quantizer implementing the Lloyd-Max algorithm \cite{Lloyd,Max}), \textcolor{black}{hardware-limited task-based quantization (HLTB) in \cite{Eldar-TSP1-2019}}  and for the proposed vector GOQ computed thanks to algorithm 1. Although Algorithm 1 is based on a HR approximation, it is seen to provide a very significant gain in terms of OL even for a small number of regions. For $M=5$ a conventional quantizer would lead to a relative OL of $70\%$ which is a significant performance degradation w.r.t. the \textcolor{black}{ideal} case where $g$ is perfectly known, whereas the proposed GOQ allows the OL to be as low as $10\%$. \textcolor{black}{Besides, compared to HLTB quantizer which is also goal-oriented, the  optimality loss reduction of proposed algorithm is still considerable in low-resolution regime.} The explanation behind this performance gain is already available through Example 1 in which we have seen the importance of adapting the ``density" or more generally the concentration of the regions (and thus allocating the quantization bits) not according to the parameter distribution (conventional \textcolor{black}{approach}) but to an appropriately weighted distribution. This difference is illustrated through Fig. \ref{fig:pdf_comparison}. The top subfigure shows the p.d.f. of the parameter $g$ (namely $\phi(g)$). The bottom subfigure shows $\lambda_{\max} \left(g; f^\text{QUA} \right) \phi \left( g\right)$. The analysis conducted in Sec. \ref{sec:scalar_approximation} suggests to concentrate the quantization regions according to this weighted density, which is markedly different from $\phi$. By doing so, Algorithm 1 provides a very significant improvement, the main powerful insight being not to allocate quantization resources to the most likely realizations of the information source but to the ones that impact the most the goal, which is measured through the weighted density $\lambda_{\max} \left(g; f^\text{QUA} \right) \phi \left( g\right)$. \textcolor{black}{Notice that the above numerical results are obtained when the p.d.f. of $g$ is known. In practice, it might happen that this p.d.f. is not available or is time-varying. Then one can easily adapt algorithm 1 by replacing statistical means with empirical/sample means and possibly, refreshing the database on the fly if the statistics need to be tracked. Fig. \ref{fig:quadratic_data_oriented} precisely shows the loss that would be induced by using a relatively small database instead of knowing the input distribution perfectly. One can observe that the data-based GO quantizer still could achieve a relative optimality loss of $9\%$ for a database with only $1000$ data points, which illustrates the relevance of the proposed method when the input distribution is not available.}

\textcolor{black}{\subsection{Goal-oriented quantization and power consumption scheduling}}

\begin{figure}[tbp]
\centering{}
\includegraphics[scale=0.45]{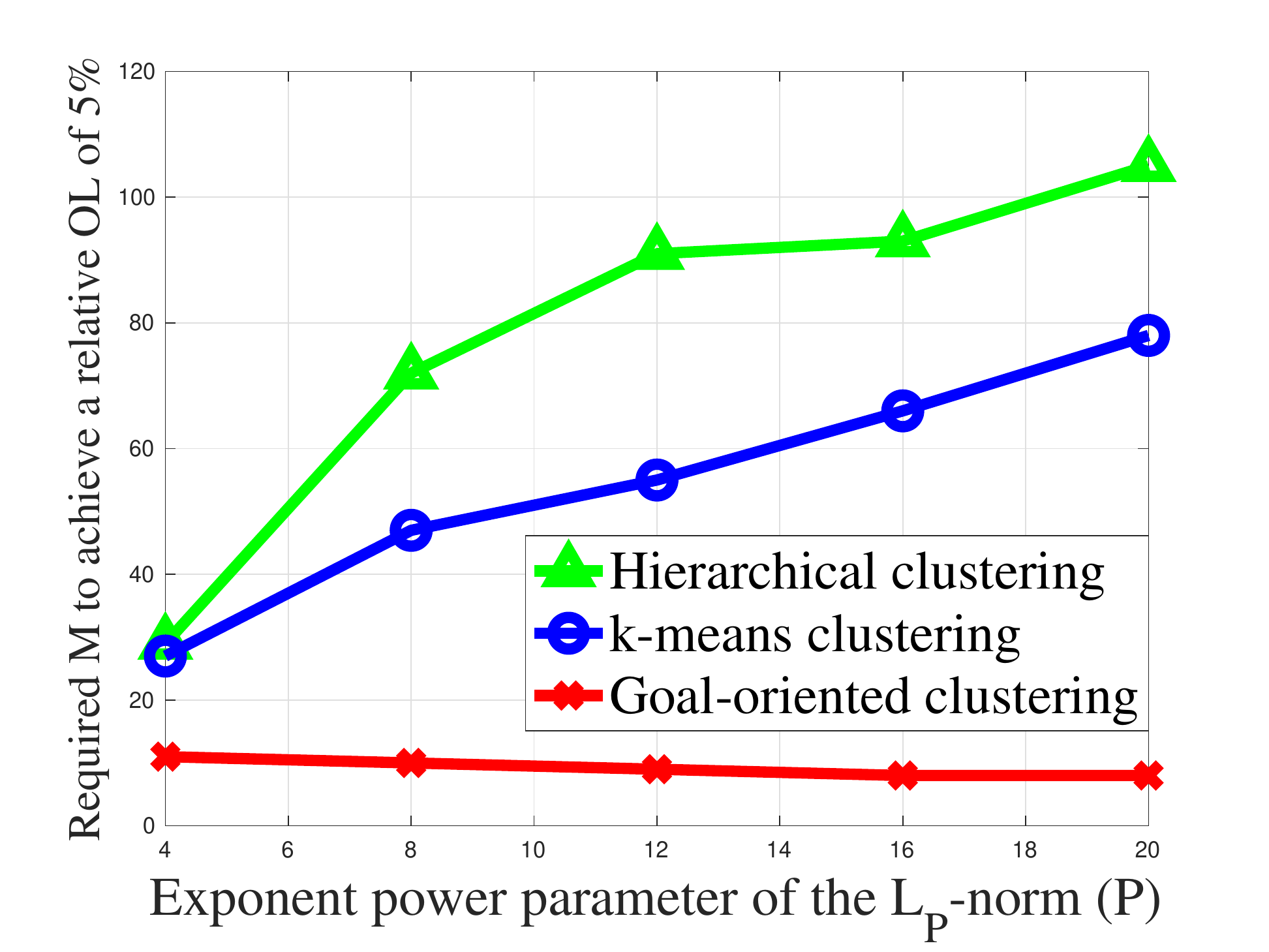}\caption{Required number of clusters ($M$) against the Exponent power parameter of the $\mathrm{L}_\mathrm{P}$-norm ($\mathrm{P}$) for the $k$-means and goal-oriented clustering. The goal-oriented clustering approach yields a drastic reduction in terms of the number of clusters when $\mathrm{P}$ increases.\label{fig:Linfty}}
\end{figure}

$\blacktriangleright$ Now we assume \textcolor{black}{$d=p=24$}. We consider a performance metric which is relevant for a communication problem in the smart grid. Indeed, we consider that the goal function $f^{\mathrm{PCS}}(x;g) = \| x+g\|_{\mathrm{P}}$, $\mathrm{P}$ being the exponent power parameter of the $\mathrm{L}_\mathrm{P}$ norm, and $\mathrm{PCS}$ stands for power consumption scheduling. This time the vector $x=(x_1,...,x_d)$ ($d=p$ here) represents the chosen flexible power consumption scheduling strategy; we impose that $ x_i \geq 0 $ and  $\sum_{i=1}^{d} x_i \geq E$, $E>0$ being the desired energy level \textcolor{black}{chosen as $30$ kWh in our simulation setting}. The parameter vector $g$ represents the non-controllable part of the power. When $\mathrm{P}$ becomes large, the problem amount to limiting the peak power. \textcolor{black}{The clustering problem is a data-based counterpart of the quantization problem in which a finite set of realizations for $g$ is available (instead of the knowledge of $\phi$).} We want to cluster a finite dataset into clusters or groups of data (instead of continuous regions). And the goal is to minimize $f^{\mathrm{PCS}}$ by only having a clustered version of the data. \textcolor{black}{For the purpose of applying the GOQ approach to clustering, we make the following two choices in terms of implementation. }First, the statistical expectation is replaced with its empirical version in the algorithm; the empirical mean is performed over the $300$ time series of the Pecanstreet dataset. Second, since the number of samples is small, representatives are computed by directly minimizing $L(\textcolor{black}{Q},f)$ (as in \cite{Zhang-AE-2021}) instead of the approximated version $\widetilde{L}$. For a given relative OL of $5\%$ one then looks at the number of required clusters (that is, $M$) versus the exponent power parameter of the $\mathrm{L}_\mathrm{P}$ (that is, $\mathrm{P}$). In Fig. \ref{fig:Linfty}, \textcolor{black}{we compare the performance of the the GO clustering technique with the $k-$means algorithm (which is exactly {the data-based counterpart} of the LM algorithm) and hierarchical clustering (HC) algorithm for the Pecanstreet database \cite{pecanstreet}. For HC, the squared Euclidean distance and weighted pair group method with arithmetic mean are used. First one can observe that partitioning clustering slightly outperforms hierarchical clustering, this might be explained by the fact that several clusters in HC compose of a single outlier data point (in terms of Euclidean distance), but outlier data points might yield similar decision as normal data points for Lp-norm problems especially with large $p$.} For $\mathrm{P}$ ranging from $4$ to $20$, the figure shows that the number of required clusters can be decreased from about $M=80$ to $M=8$ by adapting the clustering technique to the final decision instead of creating clusters based on an exogenous similarity index, which is the Euclidean norm in the case of the $k-$means algorithm.       

\section{Conclusion}
\label{sec:Conclusions}

In this paper, the focus is on one key element of a goal-oriented communication chain namely, the quantization stage. The GOQ problem is very relevant for lossy data compression e.g., to have high spectral efficiency in wireless systems (by transmitting only the minimum amount of information relevant to the correct task execution). It is also relevant for many resource allocation problems, hence the choices for the goal function in this paper. One of the contributions of this paper is to exploit the HR assumption both for the analysis and design of a GOQ. Valuable insights of practical interest have been obtained. Let us mention two of them. \textcolor{black}{The most conventional way of designing a source coder is to allocate 
resources (say bits) according to the frequency of the realization of the source symbol (this is what Huffman and arithmetic coding schemes and their many variants do). Our analysis shows that this approach may lead to a significant performance degradation and rather shows in a precise way (see e.g., Prop. III.1, Example 1, and Fig. 4) how the variation speed of the goal and decision functions should be taken into account to allocate such resources in a much more efficient way.} Our analysis also allows one to make progresses into the direction of understanding how the goal function impacts the quantizer. Both analytical and simulation results are provided to exhibit the existence of possible classes of functions which would more or less easy to be compressed. This knowledge allows the quantizer to be matched to the goal. For example, rough quantization seem to have a small impact on the task execution as far as log-type goal functions are concerned. The behavior is different for exp-type functions. This suggests for example that CSI feedback should be much finer for energy-efficient performance metrics than for spectral-efficiency metrics. It is seen that the proposed framework is rich in terms of practical insights. Nonetheless, many relevant issues are left open and would need to be explored. For instance, theoretical analysis relies on smoothness assumptions for the goal and decision functions. What would the results become for non-smooth functions? The functions are also assumed to be known. How to adapt the approach when only the realizations of these functions are available? Also a dedicated complexity analysis should be conducted. Generally, the problem of designing vector GO quantizers when the dimension increases is open. \textcolor{black}{An interesting extension of this work would also be to address the case of a non-stationary source, leading to the problem of an adaptive quantizer.} How learning techniques could be used to solve all these issues?

\appendices

\section{Proof of Proposition III.1} 
By using Taylor expansion, the optimality loss in high-resolution regime can be approximated by

\allowdisplaybreaks
{\footnotesize\begin{align}
& {L} \left( \textcolor{black}{Q};f \right) \nonumber\\
= &\alpha_f\sum_{m=1}^M \int_{\mathcal{G}_m} \left[ f \left(\chi \left(z_m \right);g \right) - f \left(\chi \left(g\right);g \right) \right] \phi \left( g\right)\mathrm{d}g\nonumber\\
\overset{(a)}{=}& \alpha_f\sum_{m=1}^M \int_{\mathcal{G}_m} (\chi \left(z_m \right) - \chi \left(g \right))^{\kappa}\frac{1}{\kappa!}\frac{\partial^{\kappa}f(x;g)}{\partial x^{\kappa}}|_{x = \chi \left(g \right)}\phi(g)\mathrm{d}g + o\left(M^{-\kappa}\right)\nonumber\\
\overset{(b)}{=}& \alpha_f\sum_{m=1}^M \int_{\mathcal{G}_m} (z_m-g)^{\kappa}\left(\frac{\mathrm{d}\chi(g)}{\mathrm{d} g}\right)^{\kappa}\frac{1}{\kappa!}\frac{\partial^{\kappa}f(\chi(g);g)}{\partial x^{\kappa}}\phi(g)\mathrm{d}g+o\left(M^{-\kappa}\right)\nonumber\\
\overset{(c)}{=}& \alpha_f \int_{\mathcal{G}} \frac{\Delta^{^{\kappa}}(g)}{(\kappa+1)2^{\kappa}}\left(\frac{\mathrm{d}\chi(g)}{\mathrm{d} g}\right)^{\kappa}\frac{1}{\kappa!}\frac{\partial^{\kappa}f(\chi(g);g)}{\partial x^{\kappa}}\phi(g)\mathrm{d}g+o\left(M^{-\kappa}\right)\nonumber\\
\overset{(d)}{=}&  \frac{\alpha_f}{(2M)^{\kappa}(\kappa+1)!}\int_{\mathcal{G}} \rho^{^{-\kappa}}(g)\left(\frac{\mathrm{d}\chi(g)}{\mathrm{d} g}\right)^{\kappa}\frac{\partial^{\kappa}f(\chi(g);g)}{\partial x^{\kappa}}\phi(g)\mathrm{d}g + o\left(M^{-\kappa}\right)\nonumber\\
\label{eq:infi_derivation}
\end{align}}
(a) corresponds to the Taylor expansion of $(f(\chi(z_m);g)-f(\chi(g);g))$ in the regime of large $M$ (infinitesimals of $M^{-\kappa}$ are not considered further); (b) follows from the fact that the higher order terms in the Taylor expansion of $(\chi(z_m)-\chi(g))$ are negligible w.r.t. the first term. (c) extends the idea of approximating mean-square error distortion in high resolution regime (see \cite{Bennett_Beld_1948,Panter_IRE_1951}) to cases with even-order $\kappa$, i.e.,

{\small\begin{equation}
\begin{split}
     &\int_{\mathcal{G}_m} (z_m-g)^{\kappa}\left(\frac{\mathrm{d}\chi(g)}{\mathrm{d} g}\right)^{\kappa}\frac{1}{\kappa!}\frac{\partial^{\kappa}f(\chi(g);g)}{\partial x^{\kappa}}\phi(g)\mathrm{d}g\\   
     \approx&\left(\frac{\mathrm{d}\chi(z_m)}{\mathrm{d} z_m}\right)^{\kappa}\frac{1}{\kappa!}\frac{\partial^{\kappa}f(\chi(z_m);z_m)}{\partial x^{\kappa}}\phi(z_m)\int_{z_m-\frac{\Delta(z_m)}{2}}^{z_m+\frac{\Delta(z_m)}{2}} (z_m-g)^{\kappa}\mathrm{d}g\\
     \approx&\left(\frac{\mathrm{d}\chi(z_m)}{\mathrm{d} z_m}\right)^{\kappa}\frac{1}{\kappa!}\frac{\partial^{\kappa}f(\chi(z_m);z_m)}{\partial x^{\kappa}}\phi(z_m)\frac{\Delta(z_m)^{^{\kappa}}}{(\kappa+1)2^{\kappa}}\Delta(z_m)\\
     \approx& \int_{\mathcal{G}_m} \frac{\Delta^{^{\kappa}}(g)}{(\kappa+1)2^{\kappa}}\left(\frac{\mathrm{d}\chi(g)}{\mathrm{d} g}\right)^{\kappa}\frac{1}{\kappa!}\frac{\partial^{\kappa}f(\chi(g);g)}{\partial x^{\kappa}}\phi(g)\mathrm{d}g;
\end{split}
\end{equation}} 
(d) follows from results on high resolution quantization referring to equation (\ref{eq:definition_representative_density}). 
After the derivation optimality loss with high-resolution quantization theory, we aim to find the optimal quantization point density to minimize the OL. We first introduce a new function called value density: 
\begin{equation}
    p(g)= \left(\frac{\mathrm{d}\chi(g)}{\mathrm{d} g}\right)^{\kappa}\frac{\partial^{\kappa}f(x;g)}{\partial x^{\kappa}}|_{x=\chi(g)}\phi(g) \geq 0.
    \label{eq:def_value_density}
\end{equation}

Then we resort to the H{\"o}lder's inequality: 
\begin{equation}
\displaystyle\int p^{\frac{1}{\kappa+1}}\leq \left(\int p\rho^{-\kappa}\right)^{\frac{1}{\kappa+1}}\left(\int \rho\right)^{\frac{\kappa}{\kappa+1}}
\end{equation}
knowing $\displaystyle \left(\int \rho\right)^{\frac{\kappa}{\kappa+1}}=1$, it can be inferred that $\displaystyle\int p\rho^{-\kappa}\geq \left(\int p^{\frac{1}{\kappa+1}}\right)^{\kappa+1}$, with equality if and only if $p\rho^{-\kappa}= C_1\rho$ with $C_1 > 0$. The optimum density function of quantization points can thus be written as:
\textcolor{black}{\begin{equation}
\rho^{\star}(g)=\frac{\left[\left(\frac{\mathrm{d}\chi(g)}{\mathrm{d} g} \right)^{\kappa}\frac{\partial^{\kappa}f(\chi\left(g\right);g)}{\partial x^{\kappa}}\phi(g) \right]^{\frac{1}{\kappa+1}}}{\displaystyle\int_\mathcal{G} \left [\left(\frac{\mathrm{d}\chi(t)}{\mathrm{d} t}\right)^{\kappa}\frac{\partial^{\kappa}f(\chi\left(t\right);t)}{\partial x^{\kappa}}\phi(t)\right ]^{\frac{1}{\kappa+1}}\mathrm{d}t}.
\label{eq:lambda_op_general_app}
\end{equation}}

By plugging the optimal density into the expression of the optimality loss,  when $M$ is large, the OL ${L} \left( \textcolor{black}{Q};f\right)$ becomes:
 {\small\begin{align}
& \underset{M\rightarrow \infty}{\lim}{{L}} \left( \textcolor{black}{Q};f\right) \nonumber \\
= &\frac{\alpha_f}{(2M)^{\kappa}(\kappa+1)!}\left(\displaystyle\int_{\mathcal{G}} \left[\left(\frac{\mathrm{d}\chi(g)}{\mathrm{d} g}\right)^{\kappa}\frac{\partial^{\kappa}f(\chi\left(g\right);g)}{\partial x^{\kappa}}\phi(g)\right]^{\frac{1}{\kappa+1}} \mathrm{d}g\right)^{\kappa+1}
\label{eq:distortion}
\end{align}}

\section{Proof of Proposition IV.1}
To facilitate the derivation, we introduce the multi-index notation in order to represent partial derivative of the goal function. The $d$-dimensional multi-index can be written as ${n} = \left(n_1,\dots,n_{d} \right)$. Its sum and factorial can be expressed as $\left|{n} \right| = \sum_{t=1}^{d} n_t$ and ${n}!=\prod_{t=1}^{d} n_t!$, respectively.  Considering the  decision variable ${x}= \left({x}_1,\dots,{x}_{d} \right)$, the partial derivative with degree ${n}$ w.r.t. ${x}$ can be expressed as $    \mathfrak{D}^{{n}}_{{x}} f  = \frac{\partial^{\left|{n} \right|}f}{\partial {x}^{n_1}_1\dots \partial {x}^{n_{d}}_{d}}$, and the multi-index power of ${x}$ can be written as ${x}^n =   \overset{d}{\underset{i=1}{\prod}} {x}_{i}^{n_i}$. 

By using the Taylor expansion for multivariate functions, the optimality loss can be rewritten as:
{\small\begin{equation}
\begin{split}
 & {{L}} \left( \textcolor{black}{Q}; f \right)\\
= &\alpha_f\sum_{m=1}^M \int_{\mathcal{G}_m} \left[f({\chi}({z}_m);{g})-f({\chi} \left({g} \right);{g})\right]\phi({g})\mathrm{d}{g}\\
{=}& \sum_{m=1}^M \left[\sum_{{n}:\left|{n} \right|\leq \kappa}\int_{\mathcal{G}_m} \frac{ \mathfrak{D}^{{n}}_{{x}} f \left({\chi} \left( {g} \right);{g} \right)}{{n}!} \left({\chi} \left( {z}_m \right) - {\chi} \left( {g} \right) \right)^{{n}}\phi({g})\mathrm{d}{g}\right.\\
& \quad\quad\quad\left. +\sum_{{\widehat{n}}:\left|{\widehat{n}} \right|= \kappa+1} \int_{\mathcal{G}_m} O\left(\left({\chi} \left( {z}_m \right) - {\chi} \left( {g} \right) \right)^{{\widehat{n}}}\right)\phi({g})\mathrm{d}{g} \right]
\end{split}
\label{eq:infi_derivation_vector_general}
\end{equation}}

Interestingly, one can note that the $\frac{ \mathfrak{D}^{{n}}_{{x}} f \left({\chi} \left( {g} \right);{g} \right)}{{n}!}$ are the components of the gradient vector of $f$ w.r.t. ${x}$ when $|{n}|=1$, and $\frac{ \mathfrak{D}^{{n}}_{{x}} f \left({\chi} \left( {g} \right);{g} \right)}{{n}!}$ are the components of the Hessian matrix of $f$ w.r.t. ${x}$ when $|{n}|=2$.  For the terms with $|{n}|\geq3$, it could be seen as the infinitesimal of the second order terms. Therefore, we could take $k=2$ and ignore the higher order terms in high resolution regime. In addition, here we consider the scenario where the optimal decision function ${\chi}(.)$ always locates in the interior of the feasible set $\mathcal{X}$, and thus each component of the gradient vector is zero, namely, $\frac{\partial f({x};{g})}{\partial x_t}|_{{x}={\chi}({g})}=0$. The optimality loss can  be approximated by:

{\scriptsize\begin{equation}
\begin{split}
 &{L}\left( \textcolor{black}{Q}; f \right)\\
 {=}& \underbrace{\sum_{m=1}^M \sum_{{n}:\left|{n} \right|=2}\int_{\mathcal{G}_m} \frac{ \mathfrak{D}^{{n}}_{{x}} f \left({\chi} \left( {g} \right){g} \right)}{{n}!} \left({\chi} \left( {z}_m \right) - {\chi} \left( {g} \right) \right)^{{n}}\phi({g})\mathrm{d}{g}}_{\widehat{L}_M(Q;f)}+o\left(M^{-\frac{2}{p}}\right) \\
 \end{split}
 \end{equation}}
 and the $\widehat{L}_M(Q;f)$ can be further simplified as
 {\scriptsize\begin{equation}
 \begin{split}
&\widehat{L}_M(Q;f)=\\
\overset{(a)}{=}& \alpha_f\sum_{m=1}^M \int_{\mathcal{G}_m} \frac{1}{2}({\chi} \left( {z}_m \right) - {\chi} \left( {g} \right))^{\mathrm{T}}\mathbf{H}_f( {\chi} \left( {g} \right);{g})({\chi} \left( {z}_m \right) - {\chi} \left( {g} \right))\phi ({g})\mathrm{d}{g}\\
\overset{(b)}{=}& \alpha_f\sum_{m=1}^M \int_{\mathcal{G}_m} \frac{1}{2}(\mathbf{J}_{{\chi}}\left( {g}\right)({z}_m-{g}))^{\mathrm{T}}\mathbf{H}_f({\chi} \left( {g} \right);{g})(\mathbf{J}_{{\chi}}\left( {g}\right)({z}_m-{g}))\phi({g})\mathrm{d}{g}\\
\overset{(c)}{=}&  \alpha_f{\sum_{m=1}^M \int_{\mathcal{G}_m} \frac{1}{2}\|{g}-{z}_m\|_2^2
{e}_m^{\mathrm{T}} \mathbf{J}_{{\chi}}^{\mathrm{T}}({g})\mathbf{H}_f({\chi} \left( {g} \right);{g}) \mathbf{J}_{{\chi}} \left( {g}\right) {e}_m \phi({g})\mathrm{d}{g}}
\end{split}
\label{eq:infi_derivation_vector}
\end{equation}}
where ${e}_m$ is defined as the normalized vector of the difference, i.e., ${e}_m=\frac{{g}-{z}_m }{\|{g}-{z}_m\|_2}$. (a) follows from the fact that the second order term in the Taylor expansion can be rewritten with matrix multiplication using Hessian matrix; (b) follows from the fact that the higher order term in the Taylor expansion of $\left({\chi}({g})-{\chi}({g}_m)\right)$ are negligible w.r.t. the first order term;  (c) can be verified by defining ${e}_m$. It is worth noting that this expression is similar to the classical vector quantization while the p.d.f. of ${g}$ is weighted by a new coefficient related to the Hessian and Jacobian of the goal function and the normalized vector ${e}_m$. 
To simplify the formula, we denote by $\mathbf{A}_{f,{\chi}} \left( {g}\right)  = \mathbf{J}_{{\chi}}^{\mathrm{T}}({g})\mathbf{H}_f({\chi} \left( {g} \right);{g}) \mathbf{J}_{{\chi}} \left({g}\right)$, then one has that:
\begin{equation}
    \widehat{L}_{\textcolor{black}{M}}\left( \textcolor{black}{Q}; f \right) = \alpha_f\sum_{m=1}^M \int_{\mathcal{G}_m} \frac{1}{2}\|{g}-{z}_m\|_2^2
{e}_m^{\mathrm{T}} \mathbf{A}_{f,{\chi}} \left( {g}\right) {e}_m \phi({g})\mathrm{d}{g}.
\label{eq:approximate_optimality_loss_vector}
\end{equation}

As the normalized vector ${e}_m$ depends both on ${g}$ and the representative ${z}_m$, the vector case can not be tackled as the scalar case. Nevertheless, we will show similar properties could be found in the vector case. To directly approximate the OL defined in (\ref{eq:infi_derivation_vector}) is complicated, we thus resort to some matrix properties to bound OL. The accuracy of our approximation depends on how we approximate \textcolor{black}{the} term ${e}_m^{\mathrm{T}} \mathbf{A}_{f,{\chi}} \left( {g}\right) {e}_m$. For a given parameter ${g}$,  maximum eigenvalue and minimum eigenvalue of matrix $\mathbf{A}_{f,{\chi}} \left( {g}\right)$ are denoted by $ \lambda_{\max} ({g};f)$ and  $ \lambda_{\min} ({g};f) \geq 0 $ respectively  since the Hessian matrix $\mathbf{H}_f({\chi} \left( {g}\right);{g})$ is  \textcolor{black}{nonnegative} definite due to optimum. Therefore,  the term ${e}_m^{\mathrm{T}} \mathbf{A}_{f,{\chi}} \left( {g}\right) {e}_m$ can be upper bounded by  $\lambda_{\max}({g};f)$ and lower bounded by $\lambda_{\min}({g};f)$.

We first study the lower bound of $\widehat{L}_{\textcolor{black}{M}}\left( \textcolor{black}{Q}; f \right)$. Similarly, we extend the notation of the point density $\rho({g} )$ to a vector case which determines the approximate fraction of representatives contained in that region. Define the normalized moment of inertia of the cell $\mathcal{G}_m$ with representative ${z}_m$ by
\begin{equation}
\mathscr{M}(\mathcal{G}_m,{z}_m)=\frac{1}{p}\frac{1}{\mathrm{vol}(\mathcal{G}_m)^{1+2/p}}\displaystyle\int_{\mathcal{G}_m}\|{g}-{z}_m\|_2^2\mathrm{d}{g},
\end{equation}
and the inertial profile $\mathfrak{m}({g})=\mathscr{M}(\mathcal{G}_m,{z}_m)$ when ${g}\in\mathcal{G}_m$, the OL can be further approximated as \cite{Gersho_TIT_1979}\cite{Gray_TIT_1998}:
{ \begin{equation}
\begin{split}
& \mathrm{L}\left( \textcolor{black}{Q}; f \right) \\
= &\alpha_f\sum_{m=1}^M \int_{\mathcal{G}_m} (f({\chi}({z}_m);{g})-f({\chi} \left( {g} \right);{g}))\phi(g)\mathrm{d}{g}\\
\overset{(a)}{\geq} &\alpha_f\sum_{m=1}^M \int_{{g}\in\mathcal{G}_m} \frac{1}{2}\|{g}-{z}_m\|_2^2\lambda_{\min}({g};f)  \textcolor{black}{\phi{(g)}}\mathrm{d}{g} \\
\overset{(b)}{=}&   \sum_{m=1}^M\frac{\alpha_f p}{2{M}^{2/p}} \frac{\mathscr{M}(\mathcal{G}_m,{z}_m)}{\rho^{2/p}({z}_m)}\lambda_{\min}({z}_m;f) \phi({z}_m)\mathrm{vol}(\mathcal{G}_m) \\
\overset{(c)}{=}&   \frac{\alpha_fp}{2{M}^{2/p}}\displaystyle\int \frac{\mathfrak{m}({g} )}{\rho^{2/p}(g)}\lambda_{\min}({g} ;f) \left( \phi({g}) \right )\mathrm{d}{g} 
\end{split}
\label{eq:infi_derivation_vector_lower_bound2}
\end{equation}} 
(a) comes from the fact that ${e}_m$ is a normalized vector; (b) uses the definition of  $\mathscr{M}(\mathcal{G}_m,{z}_m)$ and the relation $\underset{M\rightarrow\infty}{\lim}\sum_{m=1}^M\mathrm{vol}(\mathcal{G}_m)\rho({z}_m)=M$; (c) is still the definition of Riemman integral. This result can be seen as a special case of Bennett's integral (see \cite{Bennett_Beld_1948}\cite{Gray_TIT_1998}) by replacing $\phi \left( {g}\right)$ by the product $\lambda_{\min}({g};f) \phi({g})$. However, it is not known how to find the optimal inertial profile $\mathfrak{m}({g})$ and it is not even known what functions are allowable as inertial profiles. To this end, Gersho \cite{Gersho_TIT_1979} made the widely accepted hypothesis or conjecture that when $R$ is large, most regions of a $p$-dimensional quantizer aims at minimizing or nearly minimizing the mean square error are approximately congruent to some basic tessellating $p$-dimensional cell shape $\mathbb{T}_{p}$. With this conjecture, the optimal inertial profile $\mathfrak{m}({g})$  can be seen as a constant $\mu_{p}$ in high resolution case. By using the H{\"o}lder's inequality, the optimal density $\rho({g})$ that minimizes the distortion can be written as
\begin{equation}
\rho^{\star}({g})=\frac{\left(\lambda_{\min}({g};f) \phi({g})\right)^{\frac{p}{p+2}}}{\displaystyle\int_{\mathcal{G}}\left(\lambda_{\min}({t};f) \phi({t})\right)^{ \frac{p}{p+2} }\mathrm{d}{t}}
\end{equation}
resulting in the low bound of distortion in (\ref{eq:lower_bound_vec}). The same reasoning can be applied to the derivation of the proposed upper bound.

\textcolor{black}{\textbf{Remark} When the number of cells is large, one has that $\mathfrak{m}({z_m}) \approx \mathfrak{m}({g})$. Then one is able to define the inertial profile $\mathfrak{m}({g})$ for the parameter $g$. Moreover, when $M$ is large, it is observed that the optimal cells (in the sense of the distortion) are roughly congruent to some basic tessellating cell shape (Gersho's conjecture). Even if it is difficult to find the optimal $\mathfrak{m}({g})$, it could be treated as a constant by admitting Gersho's conjecture since it is normalized.}

\bibliographystyle{IEEEbib}
\bibliography{strings,refs}

\end{document}